\documentclass[11pt]{article}

\makeatletter
\@addtoreset{equation}{section}
\makeatother

\usepackage{epsfig,amsfonts}
\usepackage[fleqn]{amsmath}
\usepackage{amsthm,amssymb}
\usepackage{graphicx}
\usepackage{hhline}
\usepackage{cite}
\usepackage{upgreek}
\usepackage{wasysym}
\def\be{\begin{equation}}
\def\ee{\end{equation}}
\def\bdm{\begin{displaymath}}
\def\edm{\end{displaymath}}
\def\bea{\begin{eqnarray}}
\def\eea{\end{eqnarray}}

\def\zb{{\bar z}}
\def\wb{{\bar w}}

\def\pz{\partial_z}
\def\pzb{\partial_{\bar z}}

\newcommand{\rd}{\mbox{d}}
\newcommand{\ri}{\mbox{i}}
\newcommand{\re}{\mbox{e}}

\begin{document}
\begin{titlepage}
\begin{flushright}
RUNHETC-2012-16\\
\end{flushright}

\vspace{1.cm}

\begin{center}
\begin{LARGE}
{\bf Integrable boundary interaction\\
in 3D target space:}

{\bf the ``pillow-brane''  model}

\end{LARGE}

\vspace{1.3cm}

\begin{large}

{\bf Sergei  L. Lukyanov and
Alexander B. Zamolodchikov}
\end{large}

\vspace{1.cm}

{NHETC, Department of Physics and Astronomy\\
     Rutgers University,
     Piscataway, NJ 08855-0849, USA}\\

\vspace{.2cm}

{and}\\

\vspace{.2cm}
L.D. Landau Institute for Theoretical Physics\\
  Chernogolovka, 142432, Russia\\

\vspace{.2cm}

\vspace{.2cm}

\end{center}

\begin{center}
\centerline{\bf Abstract} \vspace{.8cm}
\parbox{13cm}{
We propose a model of boundary interaction, with three-dimensional target space,
and the boundary values of the field ${\bf X}\in \mathbb{R}^3$ constrained
to lay on a two-dimensional surface of the ``pillow'' shape. We argue that the
model is integrable, and suggest that its exact solution is described
in terms of certain linear ordinary differential equation.}
\end{center}

\vspace{.2cm}

\begin{flushleft}
\rule{3.1 in}{.007 in}\\
{August 2012}
\end{flushleft}
\vfill
\end{titlepage}
\newpage

\section{ Introduction} \label{secintro}

A class of quantum field theories in 2D space-time, in which conformal invariance
is broken only by boundary conditions, is of interest both in connection with
``brane'' states in string theories \cite{Polchinski:1996na}, and as useful models of quantum
Brownian motion \cite{Callan:1989mm}. Among such theories are
the ``brane models'', where the bulk CFT consists of a collection of $N$
free massless scalar fields ${\bf X}( z,{\bar z})$, associated with the
coordinates of $N$-dimensional ``target space'', while the interaction is introduced
through nonlinear constraints imposed at the boundary of 2D ``world sheet'':
The boundary values ${\bf X}_B$ are required to lay on a nonlinear hypersurface
${\bf\Sigma} \subset \mathbb{R}^N$ - the ``brane''.\footnote{Here we use the term ``brane''
in reference to a generic boundary constraint of this kind.
Of course, on-shell stringy brane states are associated only with conformally
invariant boundary conditions \cite{Polchinski:1996na}.}

Interaction generated by such constraints requires renormalization, and
therefore in general the shape of the hypersurface ${\bf\Sigma}$ ``flows'' under
the Renormalization Group (RG) transformations. In the weak coupling regime such
RG flow reduces to an interesting case of geometric flow - the so called mean curvature
flow (for review, see \cite{Bakas:2007qm}) - in the same way as the Ricci flow
\cite{:1982Hamilton, :2002Perelman} emerges as the
weak-coupling limit of the bulk RG flow of 2D sigma models \cite{Friedan:1980jf}.
When the
curvature of ${\bf\Sigma}$ is not small (on
the scale set by the size of the quantum fluctuations of the gradients $\partial{\bf X}$),
the brane models require non-perturbative treatment. Approach beyond the perturbation theory
exists if the model is {\it integrable}. This means that the boundary
constraint is
consistent with infinitely many commuting integrals of motion of the world-sheet theory.
Integrable models
of this kind, with two-dimensional target space, and with ${\bf\Sigma}$ being certain
curves in $\mathbb{R}^2$, were previously studied in Refs.\cite{Lukyanov:2003rt,Lukyanov:2003nj}.
In this
work we extend
analysis to three-dimensional target space, and study a model in which ${\bf\Sigma}$ is
a special surface in $\mathbb{R}^3$, of the topology of $\mathbb{S}^2$ and a shape
resembling a good quality pillow. After a good nap, we found it comfortable enough
to be named the pillow-brane.\footnote{The pillow can be regarded as deformed sphere.
Integrability of the spherical-brane  model, with ${\bf\Sigma}=\mathbb{S}^2 \subset
\mathbb{R}^3$, was previously discussed in Ref.\cite{Lukyanov:2003rt}.}

As in \cite{Lukyanov:2003rt,Lukyanov:2003nj}, we will consider the simplest nontrivial
setting, in which the bulk CFT
lives inside
the disk $|z|<R$ ($( z, {\bar z})$ are standard complex
coordinates in $2D$ Euclidean space-time) and the non-conformal interaction
takes place at the boundary $|z| = R$. The bulk CFT
involves three-component scalar field ${\bf X} = ( X , Y, Z)\in {\mathbb R}^3$, and apart
from the boundary constraint (see below), the model is described by
the Euclidean action
\bea\label{action}
{\cal A}=\frac{1}{ 4\pi}\
\int_{|z|<R} \rd^2z\  \partial_a{\bf X}\cdot \partial_a
{\bf X}+{\cal A}_B\ ,
\eea
which includes the so-called $B$-field term
\bea\label{Baction}
{\cal A}_B=-\frac{\ri}{  4\pi}\
\int_{|z|<R} \rd^2z\  \varepsilon^{ab}\
{\bf B}({\bf X})\cdot ( \partial_a{\bf X}\times \partial_b
{\bf X})\ ,
\eea
The boundary interaction is introduced mainly through the
boundary constraint
\bea\label{bconstraint}
{\bf X}_B \equiv {\bf X}|_{| z|=R}\ \in\ \  {\bf\Sigma}^2\ ,
\eea
where ${\bf\Sigma}^2$ is certain surface in $\mathbb{R}^3$, of
the shape described in detail below. We assume the field ${\bf B}$
to be solenoidal,\footnote{Otherwise the nonlinear term \eqref{Baction}
would break conformal invariance in the bulk.} i.e.,
\bea\label{sel}
{\bf \nabla}\cdot {\bf B}=0\ ,
\eea
so that the last term in \eqref{action} is in fact a part of the boundary
interaction
\bea\label{baction}
{\cal A}_B=-\frac{\ri}{ 2\pi}\ \oint_{|z|=R}\rd\tau\
{\bf A}({\bf X}_B)\cdot
{\bf e}_j\ \partial_{\tau} \eta^{j}\ ,
\eea
where ${\bf A}({\bf X})$ is the vector potential,
\bea\label{magnet}
{\bf B}={\bf \nabla}\times {\bf A}\ .
\eea
Here $\tau$ is a parameter along the boundary $|z|=R$,
$\ \eta^j = (\eta^1,\eta^2)$ are some local coordinates on
${\bf\Sigma}^2$, and
\bea\label{kisi}
{\bf e}_j=\frac{\partial {\bf X}_B}{\partial \eta^j}\ .
\eea
are two tangent vectors to ${\bf \Sigma}^2$.

As was already mentioned, generally the shape ${\bf X}_B={\bf X}_B(\eta_1,\eta_2)$
of the surface ${\bf\Sigma}^2$ ``flows'' with the RG parameter $t$ (defined here as
 $t=-\log (E)$ in terms of the normalization energy scale $E$).
The RG flow equations can be derived
perturbatively, within the loop expansion, which applies in the limit when the
curvature of the brane is small. The one-loop equations can be taken from
Ref.\cite{Leigh:1989jq}.
For our model\ \eqref{action}
they read
\bea\label{RGfl}
{\bf n}\cdot {\dot {\bf X}}_{B}-
\frac{K}{  1+B_{\rm n}^2}=0\ ,
\eea
and
\bea\label{RGflo}
\frac{1}{\sqrt{g}}\,
\frac{\partial}{ \partial \eta^i}\, \big(\sqrt{g}\, g^{ij}\, D_j\, \big)=0\ ,
\eea
where
\bea\label{qqqs}
D_j=
({\bf n}\cdot {\dot {\bf X}}_{B})\ B_j+
\frac{1}{ 1+B_{\rm n}^2}\ \frac{\partial B_{\rm n}}{ \partial \eta^j}\ .
\eea
In these equations dot signifies derivative with respect to the RG parameter $t$,
$K$ stands for is the mean (external) curvature of ${\bf \Sigma}^2$,
$g_{ij}={\bf e}_i\cdot{\bf e}_j$ is the induced metric,
$g=\det( g_{ij})$, $B_{\rm n}={\bf B}\cdot{\bf n}$ denotes the normal component
of the field ${\bf B}$, and $B_j={\bf B}\cdot {\bf e}_j.$

\begin{figure}[ht]
\centering
\includegraphics[width=10cm]{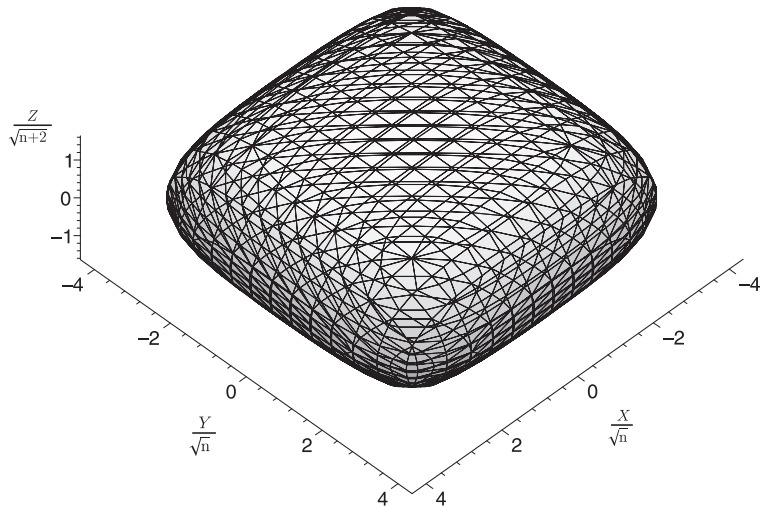}
\caption{The surface \eqref{pillow} with $n=20$, $\ \nu=\frac{1}{ 2}$ and
$\kappa^{\frac{2}{ n}}=0.01$. Its pillow-like shape gives the
name to the model. }
\label{fig-pil}
\end{figure}

In Section\ \ref{firsta} we show that with a suitable choice of the
{\it pure imaginary} field ${\bf B}$ (see Eqs.$\eqref{jddyy},\, \eqref{suuyy}$
below) the RG flow equations are satisfied by the following scale-dependent
surface:
\bea\label{pillow}
&&\sqrt{(1+w_1)(1+w_2)}\,  \cos\Big(\textstyle{\frac{Z_B}{
\sqrt{n+2}}}\Big)=\\ &&\ \ \ \
\ \ \ \ \ \ \ \ \ \ \ \ \ \sqrt{w_1 w_2}\, \cosh\Big(\textstyle{\frac{X_B}{
\sqrt{n\nu}}}\Big)+
\cosh\Big(\textstyle{\frac{Y_B}{
\sqrt{n(1-\nu)}}}\Big)
\, .\nonumber
\eea
Here $n>0$ and $0<\nu<1$ are real parameters which are independent
on the RG energy scale, while $w_{1,2}=w_{1,2}(E)$ ``flow''
with the scale. They are given by two distinct real solutions of the equation
\bea\label{sjksus}
\kappa^{2}=
w^{a_1}\  \, (1+w)^{a_3}\ \big(\,
a_2(2+ a_2)\  w^2-2a_1a_2\ w+a_1(2+a_1)\, \big)
\ ,
\eea
where
\bea\label{sksksjj}
a_1=n\nu\, ,\ \ a_2=n(1-\nu)\, ,\ \ \ a_3=-n-2\ ,
\eea
and $\kappa$ is inversely proportional to $E$, $\kappa =
\textstyle{\frac{E_{*}}{ E}}$. The
proportionality coefficient $E_{*}$ (the
integration constant of the RG flow
equation) sets the ``physical scale'' for the model:
physical quantities (like the overlap amplitudes\ \eqref{exponent}
below) will depend on the dimensionless combination $E_{*} R$.
In what follows we always take the normalization scale $E$ equal to
$R^{-1}$, so that $\kappa$ in
the left-hand side of\ \eqref{sjksus}\ coincides
with this combination,
\bea\label{kappadef}
\kappa = E_{*} R\,.
\eea
Eq.\eqref{sjksus} then relates the coefficients
$w_{1,2}$ in\ \eqref{pillow} to the radius $R$.
The shape of the surface\ \eqref{pillow} is sketched in  Fig.\ref{fig-pil}, from
which the origin of the term ``pillow-brane'' should be evident.

\noindent

Strictly speaking, the one-loop approximation applies only in the limit
$n\to\infty$, in which the curvature $K$ of the surface \eqref{pillow}
becomes uniformly small. In this limit it is superfluous to distinguish
between $n$ and $n+2$, as we do in writing  Eqs.\eqref{pillow},\,\eqref{sjksus}.
However, we believe (and will argue below) that the shape
described by these equations is ``perturbatively exact'', i.e.,
Eqs.\eqref{pillow},\,\eqref{sjksus} satisfy the RG flow equations to all
orders in the loop expansion. In this connection let us note two
symmetries of Eqs.\eqref{pillow},\eqref{sjksus}. One is the
interchange between $X$ and $Y$, or, more precisely, the transformation
\bea\label{jssyay}
X\leftrightarrow Y\, ,\ \ \  Z\to Z\, ;\   \ a_1 \leftrightarrow
a_2\, ,\ \ \ a_3\to a_3\, ; \  \ w \leftrightarrow w^{-1}\ .
\eea
The other symmetry is formally similar,
\bea\label{jddssyay}
&&X\leftrightarrow Z\, ,\ \ \ Y\to Y\, ;\    \ a_1\leftrightarrow a_3\, ,\ \ \
a_2\to a_2\, ;\ \  w
\leftrightarrow -(1+w)\ ,\nonumber\\
&&\kappa^2\to  (-1)^{a_2}\ \kappa^2\,,
\eea
but more subtle, in that it exchanges $n \leftrightarrow -(n+2)$
and therefore in effect interchanges $X$ with imaginary $Z$ in
\eqref{pillow}.

\begin{figure}[ht]
\centering
\includegraphics[width=8cm]{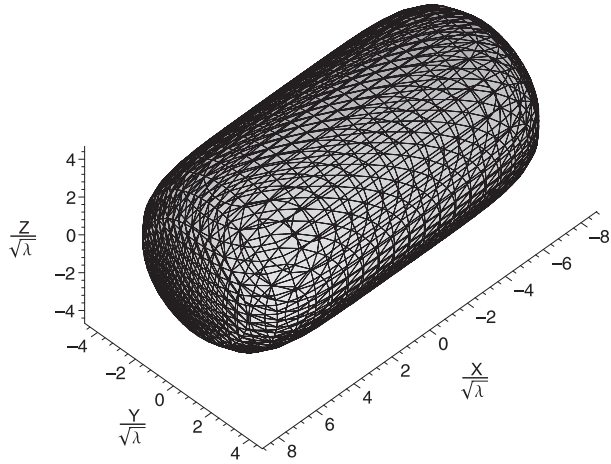}
\caption{The $U(1)$-invariant limit \eqref{nshsh} of the pillow surface, drawn with
\ ${\bar \kappa}^{\frac{2}{ \lambda}}=10^{-6}$. }
\label{fig-roll}
\end{figure}

The   general  ``pillow'' solution\ $\eqref{pillow},\, \eqref{sjksus}$
has several interesting limiting cases. Thus, when
$\nu\to 0$ and $w_1\to 0$, the pillow degenerates into the noncompact
cylindrical surface
${\bf \Sigma}^2 \to {\mathbb R}\otimes{\bf \Sigma}^1$,
where  the one-dimensional curve  ${\bf \Sigma}^1$
has a ``paperclip'' shape. That is, in this limit the first component
of the field ${\bf X}$ decouples, and
the remaining boundary QFT reduces to the paperclip-brane
model of Ref.\cite{Lukyanov:2003nj}.
Another shape can be obtained from
$\eqref{pillow},\, \eqref{sjksus}$ by taking the limit
$\nu\to 0,\ n\to\infty,\ w_{1,2}\to 0$, while
keeping the parameters $\lambda=n\nu,\ v_{1,2}= w_{1,2}
/\nu$ and ${\bar \kappa}^2=\kappa^2\, \nu^{-\lambda}\lambda^{-2}$
fixed. In this case
the pillow becomes the surface of revolution (see Fig.\ref{fig-roll}):
\bea\label{nshsh}
\frac{Y^2_B+Z^2_B}{ \lambda}=
v_1+v_2-2\sqrt{v_1 v_2}\ \cosh\Big(\frac{X_B}{\sqrt{\lambda}}\Big)
\ ,
\eea
where
$v_{1,2}$ are two real solutions of the equation:
\bea\label{skksij}
{\bar \kappa^2}=v^{\lambda}\ \re^{-\lambda v}\
 \big(\, (1-v)^2+\textstyle{\frac{2}{ \lambda}}
\, \big)\ .
\eea

Finally, taking the limit $\lambda\to\infty$,  the $U(1)$-invariant brane
\eqref{nshsh}\ turns to a sphere. Indeed, if ${\tilde \kappa^2}
={\bar\kappa^2}\lambda\, \re^{\lambda}$ is  fixed, then $v_{1,2}=1\pm \frac{1}{
\sqrt{g\lambda}}+O(\lambda^{-1})$,  where $g>0$ solves the equation,
\bea\label{aasksu} {\tilde\kappa}=\sqrt{\frac{1+2 g}{ g}}\ \re^{-\frac{1}{ 4g}}\ ,
\eea
and Eq.\eqref{skksij} reduces to
\bea\label{msshhs}
X_B^2+Y^2_B+Z^2_B=\frac{1}{ g}\ .
\eea
The sphere
$\eqref{aasksu},\, \eqref{msshhs}$ solves the RG flow equation with ${\bf
B}=0$. Some properties of this ``spherical-brane'' model was previously discussed
in Ref.\cite{Lukyanov:2003rt}.

As was already mentioned, the pillow surface\
\eqref{pillow},$\,$\eqref{sjksus} provides useful description only at $n\gg
1$, for otherwise the mean curvature at the round corners of the pillow surface
in Fig.1 is not small, and non-perturbative effects may be significant.
Moreover, even with large $n$, the perturbative  treatment is limited to sufficiently short
length scales, such that $\kappa^{\frac{2}{ n}} \ll 1$. The large length scale
behavior, where $\kappa\gg 1$, requires non-perturbative approach. In the
main body of this paper we will propose full non-perturbative description of
the pillow-brane model, valid at all scales and at all $n>0$. The description
will be given in terms of the boundary state. The boundary
state  $|\, B\, \rangle$ associated with certain boundary conditions is a
special vector in the space of states ${\cal H}$ of radial quantization of
the bulk theory, in our case
\bea\label{fullspace}
{\cal H} =
\int_{\bf P}\ {\cal F}_{\bf P} \otimes {\bar{\cal F}}_{\bf P}\,,
\eea
where
${\cal F}_{\bf P}$ is the Fock space of three-component right-moving boson
with the zero-mode momentum ${\bf P}= (P_1, P_2, P_3)$. The notion of the
boundary state is explained in \cite{Callan:1987px,Callan:1988wz} (see also \cite{Lukyanov:2003nj}), and we will
not do it here. We  just mention that its overlap
with the Fock vacuum $|\,  {\bf P}\, \rangle$  is related in a simple way to the
un-normalized one-point function of the exponential field inserted at
the center of the disk,
\bea\label{exponent}
\big\langle\,  \re^{\ri {\bf P}\cdot {\bf X}} (0,0) \,
\big\rangle_{\rm disk} =  R^{1/2-{{\bf P}^2/ 2} }\ \langle\, {\bf  P}\, |\,
B\,  \rangle\,.
\eea
This overlap amplitude $ \langle\, {\bf  P}\, |\, B\,  \rangle$,
which we sometimes call the partition function, is the main object of our interest in
this paper. When the boundary condition is not conformally invariant,
the amplitude depend on the scale parameter \eqref{kappadef}\,, and
we will use the notation
\bea\label{partit}
\langle\, {\bf  P}\, |\,
B\,  \rangle = Z(\, {\bf P}\, |\, \kappa\, )\,.
\eea

In this paper we describe some basic properties of the partition function
\eqref{partit} of the pillow brane model, and argue that this boundary
interaction is integrable. The meaning of this statement is explained in
Section\,\ref{secfive2}. Finally, we propose exact expression for the
partition function, Eq.\eqref{exactz}, in terms of solutions of linear
ordinary differential equation \eqref{diff},\,\eqref{qsaqsksai}, test it
against various expansions in the pillow-brane model, and find remarkable
agreement.

\section{Solution of one-loop RG equations} \label{firsta}

In this section, to simplify the notation we will omit the subscript
$B$ in the notation for the boundary values ${\bf X}_B$,
writing simply $(X,Y,Z)$ for $(X_B,Y_B,Z_B)$.

Let  us use  the pair of Cartesian coordinates
$(X, Y)$
as local  coordinates on the brane ${\bf\Sigma}^2$ and
write ${\bf B}=(B_X,B_Y,B_Z)$.  Then
the first RG flow equation\ \eqref{RGfl}\ becomes:
\bea\label{kjuu} L^2\ {\dot Z}=
Z_{xx}(1+Z_y^2)+Z_{yy}(1+Z_x^2)-2\, Z_xZ_y Z_{xy}\ ,
\eea
with
\bea\label{skksi}
L^2=1+Z_x^2+Z_y^2+(B_Z-Z_x\, B_X-Z_y\, B_Z)^2\ .
\eea
Here subscripts $x$ and $y$ signifies the partial derivatives
with respect to the coordinates $X$ and  $Y$, respectively.
One can check that the ansatz
\bea\label{mdjduh}
\cos\big(
\textstyle{\frac{Z}{
\sqrt{n}}}\big)=a(t)\, \cosh\big(
\textstyle{\frac{X}{
\sqrt{n\nu}}}\big)+
b(t)\, \cosh\big(\textstyle{\frac{Y}{
\sqrt{n(1-\nu)}}}\big)\, ,\eea
and
\bea\label{aasjsju}
L=\frac{\ell(t)}{ \big|\sin\big(\frac{Z}{ \sqrt{n}}\big)\big|}\, ,
\eea
with $n$ and $\nu$ being  arbitrary constants, satisfies
Eq.\eqref{kjuu}, provided the parameters $a(t), b(t)$ and $\ell(t)$,
are functions of the RG ``time'' $t$ which satisfy the following
system of ordinary differential equations:
\bea\label{lsski}
\ell^2(t)\ \nu(1-\nu)\ n\, {\dot a}&=&a\ \big(\, (1-\nu) (1-a^2)-
(1+\nu)\, b^2\, \big)\, ,\nonumber
\\
\ell^2(t)\ \nu(1-\nu)\ n\, {\dot b}&=&
b\ \big(\, \nu\, (1-b^2)+(\nu-2)\, a^2\, \big)
\ .
\eea
We further specialize  the ansatz\ \eqref{aasjsju}\  by
imposing an additional condition
\bea\label{jshsh}
\ell^2(t)=1-\frac{a^2}{\nu}-\frac{b^2}{1-\nu}\ .
\eea
With this, the form of the normal component of the $\bf B$-field is
simplified significantly. Combining
Eqs.\eqref{skksi}-\eqref{aasjsju}
 and \eqref{jshsh}\ one obtains,
\bea\label{kssi}
B_Z-Z_x\, B_X-Z_y\, B_Y=
\ri \ \, \frac{ (1-\nu)\, a\, \cosh\big(\frac{X}{\sqrt{n\nu}}\big)-\nu\, b\,
\cosh\big(\frac{Y}{\sqrt{n(1-\nu)}}\big)}{ \sqrt{\nu(1-\nu)}\
\sin\big(\frac{Z}{ \sqrt {n}}\big)}\, .
\eea

Let us turn back on the RG flow equations\ \eqref{RGflo}.
It would certainly be satisfied if one set $D_j=0$, or,  equivalently
\bea\label{kasambn}
B_j=-\frac{1}{ K}\ \frac{\partial B_{\rm n}}{ \partial \eta^j}\ .
\eea
This determines the tangential components of the field ${\bf B}$
and, with \eqref{kssi},  completely specifies the vector ${\bf B}$  at
each point
of the surface\ \eqref{mdjduh},
\bea\label{jddyy}
{B_X|}_{\bf\Sigma}&=&-\ri\ \
\frac{ a ((1-\nu)^2-b^2)\ \sinh\big({\textstyle\frac{X}{\sqrt{n\nu}}}\big)}
{\sqrt{1-\nu}\ Q}\, ,\nonumber\\
{B_Y|}_{\bf\Sigma}&=&
\ri\ \ \frac{ b\, (\nu^2-a^2)\ \sinh\big({\textstyle\frac{Y}{\sqrt{n(1-\nu)}}}\big)}
{\sqrt{\nu}\ Q}\, ,\\
{B_Z|}_{\bf\Sigma}&=&\ri\ \ \frac{ \big(b^2\nu^2-a^2(1-\nu)^2\big)\
\sin\big({\textstyle\frac{Z}{\sqrt{n}}}\big)}{
\sqrt{(1-\nu)\nu}\ Q}\, ,\nonumber
\eea
where
\bea\label{suuyy}
Q&=&a\ \big(\, (1-\nu) (a^2-1)+b^2(1+\nu)\,\big)\
\cosh\big({\textstyle\frac{X}{\sqrt{n\nu}}}\big)
+\\ &&b\ \big(\, \nu (b^2-1)+a^2(2-\nu)\,\big)\
\cosh\big({\textstyle\frac{Y}{\sqrt{n(1-\nu)}}}\big)\ .
\nonumber
\eea

At any given RG ``time'' $t$, Eqs.\eqref{jddyy} define the field ${\bf B}$ at all
points of the surface ${\bf \Sigma} = {\bf \Sigma}(t)$, Eq.\eqref{mdjduh}. However, since
the surface itself flows with $t$, these equations in fact give the field ${\bf B}$
in certain part of the bulk of $\mathbb{R}^3$, the part which is swept by ${\bf \Sigma}(t)$
in the course of the RG evolution. For this, one has to regard the components \eqref{jddyy}
as the functions of non-linear coordinates $(X,Y,t)$, with $Z$ related to these variables
through \eqref{mdjduh}. It is not difficult to verify that in this domain Eqs.\eqref{jddyy}\ indeed
define a vector which satisfies \eqref{sel}. It is straightforward to obtain
\bea\label{ddhh}
\nabla\cdot{\bf B}=\frac{\partial}{ \partial X}\, ({\dot Z}\, B_X)+
\frac{\partial}{ \partial Y}\, ({\dot Z}\, B_Y)+
\frac{\partial}{ \partial t}\, ( B_Z-Z_x\, B_X-Z_y\, B_Z)\ .
\eea
Eq.\eqref{suuyy}\ can be written in the form
$$Q=\ell^2(t)\ \sqrt{n}\, \nu(1-\nu)\ {\dot Z}\ \sin\big({\textstyle\frac{Z}{
\sqrt{n}}}\big)\ ,$$
and then the condition ${\bf \nabla}\cdot {\bf B}=0$ follows from \eqref{ddhh}.

Finally, one  has to solve the system of ordinary differential
equations\ \eqref{lsski} and \eqref{jshsh}.
Here we will restrict our attention to the case
$n>0,\  0<\nu<1$, and look for positive solutions $a, \,b >0$, such that
\bea\label{skjsusu}
a,\, b\to 0\, ,\ \ \ \ \ {\rm as}\ \ \ \ t\to-\infty\ .
\eea
Geometrically, these conditions single out compact surface
\eqref{mdjduh}, with topology of a sphere and a pillow shape shown in Fig.\,\ref{fig-pil},
which grows in the $X$ and $Y$ directions in the ultraviolet  limit $t\to -\infty$.
It is convenient to replace the positive $a$ and $b$ by parameters $w_{1,2}$ defined
through the equations
\bea\label{aajsusy}
a=\sqrt{\frac{w_1w_2}{ (1+w_1)(1+w_2)}}\, ,\ \  \
b=\frac{1}{\sqrt{(1+w_1)(1+w_2)}}\ .
\eea
We will assume that both $w_1$ and $w_2$ are real, and $w_2>w_1$, so the condition
\eqref{skjsusu}\ can be equivalently written in the form
\bea\label{smdsjsj}
w_1\to 0\, ,\ \ \ \ w_2\to+\infty\, ,\ \
\ \ \ {\rm as}\ \ \ \ t\to-\infty\ .
\eea
One can check that  both functions  $w_1(t)$ and $w_2(t)$
satisfy the same  differential equation:
\bea\label{sjsulk}
n\, {\dot w}\ \Big(\, \frac{1}{ 1+w}-\frac{\nu}{ w}\, \Big)+2=0\ ,
\eea
and hence
\bea\label{mnxg}
\re^{\frac{2}{n}(t-t_{1,2})}=\frac{ w^{\nu}_{1,2}}{1+w_{1,2}}\ ,
\eea
were $t_1$ and $t_2$ are the integration constants.
To clarify the meaning of the RG invariants $t_{1,2}$, we first
write  them as
\bea\label{ssnshh}
t_1=t_*-\frac{\mu}{ 2}\, ,\ \ \ t_2=t_*+\frac{\mu}{ 2}
\ .
\eea
Then $t_*$ can be identify with the logarithm of the physical
scale $E_*$ of the model, while $\mu$ represents
a non trivial  first integral
of the system $\eqref{lsski},\, \eqref{jshsh}$. It admits the following
representation in terms of the original coefficients $a$ and $b$,
\bea\label{djddu}
\re^{\frac{2\mu}{ n}}&=&\frac{1+b^2-a^2+\sqrt{((a-b)^2-1)((a+b)^2-1)}}{
1+b^2-a^2-\sqrt{((a-b)^2-1)((a+b)^2-1)}}\times\\ &&
\Bigg[\, \frac{1-a^2-b^2-\sqrt{((a-b)^2-1)((a+b)^2-1)}}{
1-a^2-b^2+\sqrt{((a-b)^2-1)((a+b)^2-1)}}\, \Bigg]^{\nu}\  .\nonumber
\eea
One can show that up to a simple factor, the RG invariant $\mu$ coincides
with  a total  flux of the field ${\bf B}$ \ \eqref{jddyy} through the pillow
surface,
\bea\label{kdsksi}
\mu=\frac{1}{ 4\pi\ri }\ \oiint_{\rm pillow}
\rd\eta_{1}\wedge\rd\eta_{2}\, \sqrt{g}\ B_{\rm n}\ .
\eea

Strictly speaking, presence of a nonzero flux \eqref{kdsksi} contradicts our
assumption about the solenoidal form of the field ${\bf B}$, Eq.\eqref{magnet},
and for this reason in this work we will consider only solutions with
\bea\label{ccvsjsu}
\mu=0\ .
\eea
However, we do not think that this is the last word about the flux $\mu$
in the brane models of this type. It is plausible that generalizations of the
pillow-brane  model to the case of nonzero flux is possible, and we hope to
return to this question in the future.

Once \eqref{ccvsjsu} is assumed, the parameters  $w_1$ and $w_2$ in
\eqref{aajsusy}\ become two real solutions of the same
equation
\bea\label{sulk}
\kappa^{\frac{2}{ n}}=\frac{ w^{\nu}}{ 1+w}\ ,
\eea
where $\log(\kappa)=t-t_*$. Note that this equation is the one-loop
approximation to \eqref{sjksus}, which reduces to \eqref{sulk} in the limit
$n\to\infty$. We believe that \eqref{sjksus} incorporates all higher-loop
corrections.

\section{\label{secthree2} Semiclassical analysis of the pillow-brane model}

The one-point function\ \eqref{exponent}
admits representation in terms of the functional integral,
\bea\label{path}
\big\langle\,  \re^{  \rm{i}\, {\bf P}\cdot {\bf X}} (0,0)
\, \big\rangle_{\rm disk}=\int\,
{\cal D}X\,{\cal D}Y{\cal D}Z\  \re^{ { \rm{i}}  P_1X +
 {\rm{i}} P_2Y+ {\rm{i}} P_3Z}(0,0)\ \re^{-{\cal A}[X,Y,Z]}\ ,
\eea
where ${\bf P}=(P_1,P_2,P_3)$, and
the integration is over the fields $X(z,{\bar z})$,\ $Y(z,{\bar z})$ and
$ Z (z,{\bar z})$, subjects to the constraint \ \eqref{pillow} at the boundary
$|z|=R$. It will be useful to regard the components $P_j$ of the zero mode momentum
${\bf P}$ as complex variables. Since the pillow surface is compact, the integrand
\eqref{path} is bounded for any complex vectors ${\bf P}$, hence the
overlap\ $\langle\,  {\bf P}\, |\,  B\,  \rangle$\ in Eq.\eqref{exponent}
is an entire function of its components. When $P_j$  are taken to be pure imaginary,
these parameters can be interpreted as external fields coupled to the boundary values
${\bf X}_B=(X_B,Y_B,Z_B)$. One makes a shift
\bea\label{shift}
{\bf X}\to {\bf X}+\ri\, {\bf P}\ \log {\textstyle\frac{{|z|}}{ {R}}}\, ,
\eea
of the integration variables in \eqref{path}, bringing it to the form
\bea\label{cyl}
\big\langle\,  \re^{ \rm{i}\, {\bf P}\cdot {\bf X}} (0,0)
\, \big\rangle_{\rm disk}=
R^{-{\bf P}^2/2 }\
\int\,{\cal D}{\bf X}\ \re^{-{\cal A}[{\bf X}]-{\cal
B}[{\bf X}_{B}]}\ ,
\eea
where
\bea\label{Bction}
{\cal B}[{\bf X}_{B}] = -  \oint_{|z|=R}\frac{ \rd z}{ 2\pi
z}\ \, \big( P_1X_{B}+P_2Y_{B}+P_3Z_B \big) (z)\, .
\eea

The representation\ \eqref{cyl} is most
useful in the semiclassical limit. The semiclassical approximation is valid when the
curvature of the pillow surface \ \eqref{pillow}\  is uniformly small. For this, one needs
to have $n\gg 1$ and also sufficiently small $\kappa$, such that $\kappa^{1\over n} \lesssim 1$.
Note that according to\ \eqref{sjksus}\ one has to have sufficiently
small $R$ in order to meet the last condition. Therefore, semiclassical regime in the
pillow-brane model corresponds to large $n$ and sufficiently small
length scales. In the leading semiclassical approximation the
contribution in the path integral   is dominated by the
classical solutions minimizing ${\cal A}[{\bf X}]+{\cal
B}[{\bf X}_{B}]$.

Let us first assume that the parameters $P_j$ are  small, so
that the effect of the boundary term\ \eqref{Bction}\
saddle-point configurations is negligible. We write
\bea\label{kl}
(P_1,P_2,P_3) ={\textstyle \frac{2}{\sqrt{n}}}
 \ \big(\, {\textstyle {\alpha\over \sqrt{\nu}},\, {\beta\over \sqrt{1-\nu}},\,
\gamma\, }\big)
\eea
and assume that $\alpha,\, \beta,\, \gamma$ remain finite in the limit $n\to\infty$.
The action $\mathcal{A}[{\bf X}]$ is minimized by trivial classical
solutions - the constant fields ${\bf X}(z,{\bar z}) = {\bf X}_0$, where
${\bf X}_0 =(X_0 , Y_0,Z_0)$ is any point  on the pillow surface \ \eqref{pillow}. The classical
limit of \ \eqref{partit} can be written as the integral
\bea\label{mini}
Z_{\rm class} = \oiint_{ \rm pillow}
\rd{\cal M}({\bf X}_0)\ \re^{2{\rm i}\,  \big(\, {\alpha {X_0}\over \sqrt{n\nu}} +
{\beta Y_0\over\sqrt{n(1-\nu)}}+{\gamma Z_0\over \sqrt{n}}\,  \big)}\ ,
\eea
The integration measure $\rd{\cal M}({\bf X}_0)$ is determined by
integrating out Gaussian fluctuations around the classical solution. In the presence
of the $B$-field  the  measure was calculated in Ref.\cite{Fradkin:1985qd}.
Up to the constant factor it has the Dirac-Born-Infeld form
\bea\label{mdsdjhjd}
\rd{\cal M}={\rm g}_D^3
\ \sqrt{g(1+B_n^2)}\ \ \ {\rd\eta_1\wedge \rd \eta_2\over
(2\pi)^2}\  ,
\eea
where
${\rm g}_D = 2^{-{1\over 4}}$ is  the boundary degeneracy
associated with
the Dirichlet conformal boundary condition.\footnote{The definition is as
follows: ${\rm g}_D = \langle\,  P\, |\, B_D\, \rangle$,
where $|\,B_D\,\rangle$ is
the boundary state of {\it uncompactified} boson $X$ with the
Dirichlet boundary
condition $X_B=0$, and the primary states $|\, P\,  \rangle$ are
delta-normalized, $\langle\,P\,\mid \,P'\,\rangle = \delta(P -P')$.}
In addition, the Gaussian fluctuations give rise to the one-loop term in
the renormalization of the  parameters $w_1$, $w_2$ in \eqref{pillow},
as is described by Eq.\eqref{sulk}.

Using $(X,Y)$ as local coordinates on the pillow surface, the measure
\eqref{mdsdjhjd}\ can be written as
\bea\label{ssjsu}
\rd{\cal M}({\bf X}_0)={\rm g}_D^3\ \  {\ell(t)\over (2\pi)^2}\
{\rd X_0\wedge \rd Y_0\over \big|\sin\big({Z_0\over\sqrt{n}}\big)\big|}\ ,
\eea
where $\ell(t)$ is given by \eqref{jshsh}  and \eqref{aajsusy}.
Therefore the semiclassical partition function \eqref{mini}\ admits the following
representation
\bea\label{hssy}
Z_{\rm class} = {n\, {\rm g}_D^3\over 2\pi}\
\sqrt{(\nu-(1-\nu)w_1)((1-\nu)w_2-\nu)}\ \
{\cal I}(\alpha,\, \beta, \gamma)\, ,
\eea
where
\bea\label{skss}
&&{\cal I}(\alpha,\beta,\gamma)={1\over 2\pi}\
\int_{-\infty}^{\infty}\rd x
\int_{-\infty}^{\infty}\rd y
\int_{-{\pi\over 2}}^{\pi\over 2}\rd z
\  \re^{2 {\rm{i}} (\alpha x+\beta y+\gamma z)}\times \\ &&
\ \ \ \ \ \delta\big(\sqrt{w_1w_2}\,
\cosh x+\cosh y-\sqrt{(1+w_1)(1+w_2)}\, \cos z\big)\ .
\nonumber \eea
The integral in\ \eqref{skss}\ is calculated in closed form, in terms of the
hypergeometric function,
\bea\label{sjsjsu}
{\cal I}(\alpha,\beta,\gamma)=
{ Q_{-\alpha,\beta,\gamma}(w_2) \, Q_{\alpha,\beta,\gamma}(w_1)-Q_{\alpha,
\beta,\gamma}(w_2)\,
Q_{-\alpha,\beta,\gamma}( w_1)\over 2 \ri\, \alpha \sqrt{(1+w_1)(1+w_2)}}\ ,
\eea
where
\bea\label{nshshaa}
Q_{\alpha,\beta,\gamma}(w)&=&w^{-\rm{i} \alpha}\, (1+w)^{{1\over 2}
-\gamma}\times \\
&&{}_2F_{1}
\big({\textstyle{1\over 2}}-\ri\, \alpha+\ri\, \beta
-\gamma,\,
 {\textstyle{1\over 2}}-\ri\, \alpha-\ri\, \beta-\gamma,\, 1-
2\,\ri\, \alpha;-w\big)\, .\nonumber
\eea
Alternative form of \eqref{hssy}\ is obtained by transforming to the hypergeometric
functions of the argument $w^{-1}$,
\bea\label{kjsjsu}
Z_{\rm class} = \sum_{\varepsilon,\varepsilon'=\pm 1}B_{\rm class}(\, \varepsilon\, \alpha,\,
\varepsilon'\, \beta,\, \varepsilon\varepsilon'\, \gamma\, )\ F_{\rm class}
(\, \varepsilon\,  \alpha,\,  \varepsilon'\, \beta,\, \varepsilon\varepsilon'\, \gamma\,  |\,  \kappa\, )\, ,
\eea
where
\bea\label{ssiusau}
B_{\rm class}( \alpha,  \beta,\gamma)=
 {n\,{\rm g}^3_D\over 2\pi}\
{ \sqrt{ \nu(1-\nu)}\
 w_1^{\rm{ i} \alpha}\, w_2^{-\rm{i} \beta}\
\Gamma(-2\,\ri\, \alpha)\, \Gamma(-2\,\ri\,\beta)
\over \Gamma({1\over 2}-\ri\, \alpha-\ri\, \beta -\gamma)
\Gamma({1\over 2}-\ri\, \alpha-\ri\, \beta +\gamma)}
\eea
and
\bea\label{shsyy}
F_{\rm class}(\alpha,\beta,\gamma\,|\,\kappa)&=&
{
\sqrt{{(\nu-(1-\nu)w_1)((1-\nu)w_2-\nu)\over \nu(1-\nu)(1+w_1)(1+w_2)}}
}\times\\ &&
w_1^{-\rm{i}\alpha} w_2^{\rm{i}\beta}\ Q_{-\alpha,\beta,\gamma}( w_1)\,
Q_{-\beta,\alpha,\gamma}(w_2^{-1})\ .
\nonumber
\eea
Advantage of this representation is that it makes explicit the singular behavior of
$Z_\text{class}$  at short scales $\kappa \to 0$, since in this limit we have
\bea\label{siais}
w_1\sim  \kappa^{2\over n\nu}
\to 0\, ,\ \ \ w_2\sim  \kappa^{-{2\over n(1-\nu)}}\to
\infty\ .
\eea

The above result was derived under the assumption that
the zero-mode momenta $P_j$ are small. But it is not too difficult to extend it
to much larger values of these parameters. When $P_j$ become comparable to
$\sqrt{n}$, the vertex insertion in\ \eqref{path}\ must be treated as a  part
of the action, as  it affects the saddle-point configurations.
The saddle-point configuration(s) is still a constant field,
${\bf X}(z,{\bar z}) = {\bf X}_0$, but now ${\bf X}_0$ is not an arbitrary point on
the surface\ \eqref{pillow},
but has to extremize the boundary action \eqref{Bction}, which for constant
fields takes the form
\bea\label{jsxs}
{\cal B}[\, {\bf X}_0\, ]= -\ri\, {\bf P}\cdot{\bf X}_0\ .
\eea
Therefore, the saddle-points are the points where the tangent plane to the pillow surface
is perpendicular to the vector ${\bf P}$. The dominating saddle point is easier to identify
in the case of pure imaginary ${\bf P}$: it is the point of the real-space pillow surface
farthest in the direction of $-\ri\,{\bf P}$. Contribution of this point can be determined as follows.
The saddle-point action, together with the Gaussian integral over the constant mode can be
simply taken from the asymptotic $\alpha,\, \beta,\, \gamma\to -\ri\, \infty$ of the expression
\eqref{hssy}, since the asymptotic of the integral  \eqref{skss} is controlled by the very
same saddle point. However, when $(P_1,P_2,P_3) \sim \sqrt{n}$ the Gaussian integrals over the
non-constant modes can not be ignored.
For small deviations from the saddle point ${\bf X}_0$, let us write
\bea\label{ksxusiu}
{\bf X}(z,{\bar z}) - {\bf X}_0 = {\bf e}_j({\bf X}_0)\ \xi^j(z,{\bar z}) +
{\bf n}({\bf X}_0)\ \delta X_{\perp} (z,{\bar z})\ ,
\eea
where we assume that the vectors ${\bf \re}_j({\bf X_0})$ tangent to the pillow surface
at ${\bf X}_0$, together with the normal vector ${\bf n}({\bf X}_0)$, form orthonormal
basis in $\mathbb{R}^3$. Furthermore, we choose ${\bf e}_j({\bf X_0})$ in such a way
that in the quadratic approximation
\bea\label{uyjsujs}
\Big[\, \delta X_{\perp}+K_1^{(0)}\ {\textstyle{(\xi^1)^2\over 2}}+
K_2^{(0)}\ {\textstyle {(\xi^2)^2\over 2}}\, \Big]\Big|_{B}=0\  ,
\eea
where $K_j^{(0)}$ are principal curvatures of the pillow at
${\bf X}_0$. Then the Gaussian term in the full boundary action takes the form
\bea\label{ksisia}
{\cal A}_B+{\cal B}=
\oint_{|z|=R} {\rd\tau\over 2\pi}\,
\Big[\,
{\textstyle{m_1\over 2R}}\, (\xi^1)^2+{\textstyle{m_2\over 2R}}
\, (\xi^1)^2-\ri\
  B^{(0)}_{\rm n}\ \varepsilon_{ij}\, \xi^i\partial_\tau\xi^j
\, \Big]\, ,
 \eea
where
\bea\label{kssusu}
m_j=-|-\ri\, {\bf P}|\ K^{(0)}_j\ ,
\eea
and $B^{(0)}_{\rm n}$ the normal component of ${\bf B}({\bf X}_0)$.

We see that, while to the leading approximation the normal component $\delta
X_{\perp}$ of the field ${\bf X}$ still can be treated with the Dirichlet
boundary condition, the tangential components $\delta {\bf X}$ have free
boundary conditions with the ``boundary mass terms'' in \eqref{ksisia}.
Note that for $m_j \sim 1$ (the condition which we assume) the
energy scale associated with this ``boundary mass'' is $\sim R^{-1}$,
so that the use of the renormalized parameters $w$ defined as in\ \eqref{sulk}
is still legitimate. The boundary amplitude of the free field with quadratic boundary
interaction is well known (see \cite{Witten:1992cr, Andreev:2000yn}). One finds that
Eq.\eqref{kjsjsu} would apply to the case of $(P_1,P_2,P_3)\sim \sqrt{n}$ as well
if one simply replaces  $B_{\rm class}=B_{\rm class}(\alpha,\beta,\gamma)$  in
\ \eqref{kjsjsu} by
\bea\label{shsdy}
{\tilde B}_{\rm class} =
{B}_{\rm class}\Big(\textstyle{\sqrt{n\nu}
\over 2}\, P_1 , \textstyle{\sqrt{n(1-\nu)}\over 2}\, P_2, \textstyle{\sqrt{n}\over 2}\,
P_3\Big)\,
\Gamma\big(1 -{\textstyle {\ri\, \Pi_1\over \sqrt{n}}}\big)\,
\Gamma\big(1 -{\textstyle {\ri\, \Pi_2\over \sqrt{n}}}\big)\,  ,
\eea
where
\bea\label{nash}
\Pi_1&=&{1+w_1\over \nu-(1-\nu)\, w_1}\ \sqrt{U(w_1)}\\
\Pi_2&=&{1+w_2\over (1-\nu)\, w_2-\nu}\ \sqrt{U(w_2)}
\ , \nonumber
\eea
with
\bea\label{asxaaa}
U(w)={\nu\,  P_1^2+w\ (1-\nu)\, P_2^2\over 1+w}+
{w\ P_3^2\over (1+w)^2}\ .
\eea
In Eq.\eqref{nash}\ the branch of the square root  should be chosen in such a
way that $\Im m \, \Pi_{1,2}>0$ for pure imaginary $P_{1,2}$ with
$\Im m \, P_{1,2}>0$.
Of course in this case the arguments of $B_\text{class}$ in \ \eqref{shsdy} are large,
and one can use the asymptotic forms of \eqref{ssiusau}
and \eqref{shsyy}.

\section{\label{secfour2} Ultraviolet limit of the pillow-brane model}

\begin{figure}[ht]
\centering
\includegraphics[width=10cm]{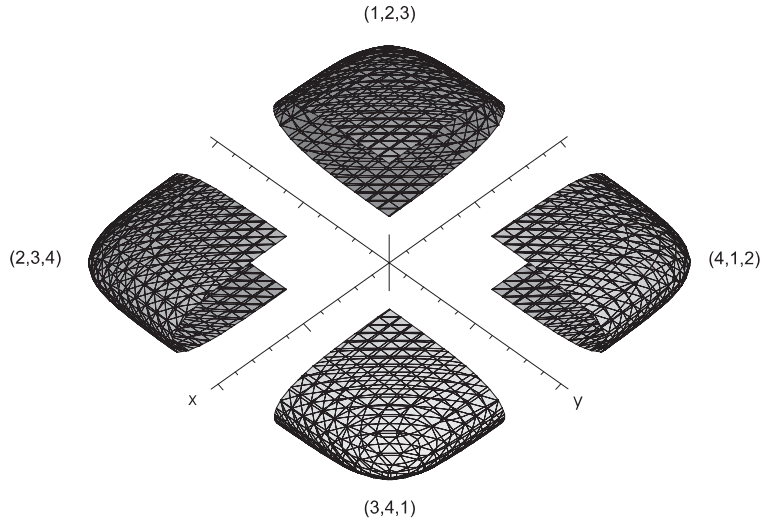}
\caption{The pillow viewed as the composition of four "corners".}
\label{fig-corn}
\end{figure}

\begin{figure}[ht]
\centering
\includegraphics[width=7cm]{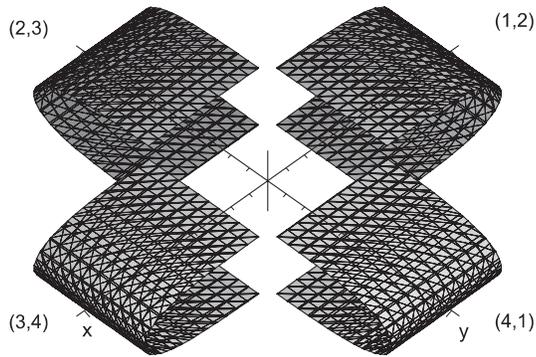}
\caption{Four hairpin cylinders}
\label{fig-cross}
\end{figure}

As was mentioned in  Introduction, at short length scales the
$w_1$ and $w_2$  in \eqref{pillow} tend to $0$ and to $\infty$, respectively.
Correspondingly, the pillow surface grows wide in the $X$ and $Y$ directions,
while its size in the $Z$ direction approaches $\pi\,\sqrt{n+2}$. In this
limit the pillow can be regarded as a juxtaposition of four ``corners'', as is
shown in Fig.\,\ref{fig-corn}.
Any two adjacent corners are connected to each other  via
the "hairpin cylinder". By this term we understand
intrinsically flat surface obtained by tensoring $\mathbb{R}$ and the ``hairpin curve''
defined in Ref.\cite{Lukyanov:2003nj}. There are four hairpin cylinders in the UV
limiting form of the pillow, which we label $(1,2)$, $(2,3)$, $(3,4)$, and $(4,1)$, as
shown in Fig.\,\ref{fig-cross}.

When the pillow is wide, and if $\Im m \,P_{1}$ and $\Im m \,P_{2}$ are not too
small, the functional integral in \eqref{cyl} is dominated by fields localized near
one of the corners in Fig.\,\ref{fig-corn}. Which one of the four corners contributes
depends on the signs of $\Im m\,P_{1}$ and $\Im m\,P_{2}$. It is useful therefore to
describe the boundary states associated with the ``corner'' and ``hairpin cylinder''
branes in some details.

\subsection{Hairpin cylinder}

As the hairpin cylinder surface is a direct product of a line and the hairpin curve,
all basic properties of the associated boundary state can be
taken from Ref.\cite{Lukyanov:2003nj}. Let us consider first  the hairpin
cylinder $(1,2)$. This surface is defined by the equation
\bea\label{papercyl}
\cos\Big({\textstyle{Z_B\over
\sqrt{n+2}}}\Big)=
 {{\tilde a}}\ \exp\Big(-{\textstyle{X_B\over
\sqrt{n\nu}}}\Big)
\, ,
\eea
where
\bea\label{RGtimepap}
{\tilde a} =\kappa^{1\over n\nu}\, ,
\eea
and  $\kappa=E_*R$, while the field ${\bf B}$ in this case is a constant
vector pointing in the $X$ direction,
\bea\label{ssuasuy}
{\bf B}=\ri\ (-\sqrt{1-\nu},\, 0,\, 0)\ .
\eea
The constraint \eqref{papercyl} does not involve $Y_B$, i.e., the field $Y$
obeys the free (von Neumann) boundary condition.

The RG flow just shifts the hairpin cylinder homogeneously in the $X$
direction, and therefore this boundary condition is equivalent to a fixed
point of the RG transformation. This fact is made explicit by redefining
the RG transformation by supplementing it with the field redefinition
$(X,Y,Z)\to(X + \frac{\delta t}{\sqrt{n\nu}}, Y, Z)$. This
corresponds to introducing the linear dilaton $\Phi({\bf X}) =-{X\over \sqrt{n\nu}}$,
or, which is much the same, adding the ``improvement'' term to the energy-momentum tensor,
\bea\label{energymom}
T(z)&=& - \partial_z  {\bf X}\cdot \partial_z {\bf X}
-{1\over \sqrt{n\nu}}\
\partial_{z}^2{ X}
\,,\\ \nonumber
{\bar T}({\bar z}) &=& - \partial_{\bar z}
{\bf X}\cdot \partial_{\bar z} {\bf X}
- {1\over \sqrt{n\nu}}\
\partial_{\bar z}^2{ X}\, .
\eea
With this, the boundary state
$|\, B\, \rangle_{(1,2)}$ associated with the
hairpin cylinder brane enjoys conformal invariance in the
usual form,
\bea\label{cftinvaa}
\big[\,
z^2\ T(z) - {\bar z}^2 \ {\bar T}({\bar z})\, \big]_{|z|=R} \
|\, B\, \rangle_{(1,2)} = 0\,.
\eea

The hairpin cylinder brane has an extended conformal symmetry (the
W-algebra). In fact, there are many ways to introduce the W-algebra in this
theory, but only one is useful for our purposes here. We will call it
$\mathcal{W}^{(1,2)}$-algebra, with the superscript $(1,2)$ placed as the reminder
that it belongs to the hairpin cylinder $(1,2)$. The algebra $\mathcal{W}^{(1,2)}$
is generated by holomorphic currents $W_s$ of spin $s$ characterized by the condition
that they commute with two ``screening charges'',
\bea\label{skanzzb}
\oint_z\
\rd w\ W_s (z)\,\,
\re^{
{\boldsymbol
\alpha}_j
\cdot{\bf X}_R}(w)  =0\,
\eea
for $j=1,2$, where
\bea\label{screenaa}
{\boldsymbol \alpha}_1&=&
\big(\, -\sqrt{n\nu}\, ,\,  -\sqrt{n(1-\nu)}\, ,\, \ri\, \sqrt{n+2}\, \big)\,
\,\\
{\boldsymbol \alpha}_2&=&\big(\, -\sqrt{n\nu}\, ,\,  \sqrt{n(1-\nu)}\, ,-
\ri\, \sqrt{n+2}\, \big)\, .\nonumber
\eea
The integration in Eq.\eqref{skanzzb}
is taken  over a small contour around the point $z$.
The subscript $R$ stands for
the holomorphic part of the local field
${\bf X}(z,{\bar z})={\bf X}_R(z)+{\bf X}_L({\bar z})$.
The full W-algebra $\mathcal{W}^{(1,2)}$ can be generated by OPE of three
basic currents, the spin-2 energy-momentum tensor $W_2\equiv T$
(given by \ \eqref{energymom}), and the spin-1 and spin-3 currents
\bea\label{ksamksaj}
W_1(z)&=&\ri\ \sqrt{n+2}\ \partial_z Y+\
\sqrt{n(1-\nu)}\ \partial_z Z\, ,\\
W_3(z) &=& {3n\nu+2\over 3}\ \big(\partial_z {\hat Z}\big)^3 +
n\nu\ \big(\partial_z X\big)^2 \partial_z {\hat Z} -\nonumber\\
&& {n\nu\sqrt{n\nu}\over 2}\ \partial^2_z
X\partial_z {\hat Z}
+ {(n\nu+2)\sqrt{n\nu}\over 2}\ \partial X \partial^2_z {\hat Z} +
{{n\nu+2}\over 12}\ \partial^3_z {\hat Z}\ ,
\nonumber
\eea
where
\bea\label{sshyys}
\partial_z{\hat Z}=\sqrt{n+2\over n\nu+2}\ \partial_z Z+
\ri\ \sqrt{n(1-\nu)\over n\nu+2}\ \partial_z Y\ .
\eea
the higher currents $W_s$ can be found either by a direct
computation of the OPE with the screening
exponentials\ \eqref{screenaa} (the condition \ \eqref{skanzzb}
is equivalent to the statement that the singular part of the
OPE of $W_s (z)\, \re^{{\boldsymbol \alpha}_{j}\cdot
{\bf X}_R} (w)$ is a total
derivative $\partial_w (\ldots )$\,),
or recursively, from the singular parts of
the OPE of the lower currents, starting with
$W_1,\, W_2$ and $W_3$. Note that although the component $Y$ of the
field ${\bf X}$ largely plays the role of a spectator in the dynamics
of the hairpin cylinder $(1,2)$, this component is mixed in a nontrivial
way in the currents $W_s$.

There is of course an antiholomorphic counterpart of the W-algebra
$\mathcal{W}^{(1,2)}$, defined by the relations \ \eqref{skanzzb} with
$\zb, \wb$ instead of $z,w$, ${\bf X}_R (z)$ replaced by ${\bf X}_L(\zb)$,
and ${\boldsymbol \alpha}_1$, ${\boldsymbol \alpha}_2$ changed to
\bea\label{scrna}
{\bar {\boldsymbol \alpha}}_1&=&
\big(\, -\sqrt{n\nu}\, ,\,  -\sqrt{n(1-\nu)}\, ,\,- \ri\, \sqrt{n+2}\, \big)\, ,
\\
{\bar {\boldsymbol \alpha}}_2&=&\big(\, -\sqrt{n\nu}\, ,
\,  \ \sqrt{n(1-\nu)}\, ,\,
\ \ri\, \sqrt{n+2}\, \big)\, .\nonumber
\eea
The antiholomorphic currents ${\bar W}_s(\zb)$ can be obtained from $W_s(z)$
by replacing $\pz X,\, \pz Y$ $\to$ $\pzb X,\, \pzb Y$, and $\pz Z$ $\to$ $ -\,\pzb Z$,
for instance
\bea\label{snasah}
{\bar W}_1({\bar z})=\ri\ \sqrt{n+2}\ \partial_{\bar z} Y-\
\sqrt{n(1-\nu)}\ \partial_{\bar z} Z\, .
\eea
The W-algebra symmetry of the boundary state associated with the hairpin cylinder
$(1,2)$ is expressed as the set of local conditions
\bea\label{wcftinva}
\big[\,
z^s\ W_s(z) -(-1)^{s}\
{\bar z}^s \ {\bar W}_s({\bar z})\, \big]_{|z|=R} \
|\, B\, \rangle_{(1,2)} = 0\, ,
\eea
for all the W-currents in $\mathcal{W}^{(1,2)}$. At present, the status of this
statement is as follows. We have verified directly that it
is true in the classical limit $n\to \infty$, where the calculation
amounts to checking that the differences $z^s\ W_s(z) -(-1)^{s}\
{\bar z}^s \ {\bar W}_s({\bar z}),\ s=1,\,2,\,3$
vanish at the boundary $|z|=R$ in virtue of the classical equations
corresponding to the action \eqref{action},\,\eqref{Baction},
\bea\label{sksasaaa}
\Big( \partial_{\tau}{\bf X}-{\ri\over B_n}\ {\bf n}\times \partial_{\sigma}
{\bf X}
\Big)\Big|_{|z|=R}=0\, ,
\eea
together with the constraint \eqref{papercyl}. In Eq.\eqref{sksasaaa}\
$\partial_{\tau}X$ and $\partial_{\sigma}{\bf X}$ stand for
the tangential and internal
normal derivative to the boundary circle $|z|=R$,
respectively.

Other hairpin cylinders in Fig.\,\ref{fig-cross} can be described in a similar way, with
obvious modifications. The associated W-algebras are defined as follows. Introduce
two additional vectors
\bea\label{eenaa}
{\boldsymbol \alpha}_3&=&\big(\,
\sqrt{n\nu}\, ,\,  -\sqrt{n(1-\nu)}\, ,\,
-\ri\, \sqrt{n+2}\, \big)\, ,\\
{\boldsymbol \alpha}_4&=&\big( \,
\sqrt{n\nu}\, ,   \ \sqrt{n(1-\nu)}\, ,\,
\ \ri\, \sqrt{n+2}\, \big)\, .\nonumber
\eea
The W-algebras $\mathcal{W}^{(2,3)}$, $\mathcal{W}^{(3,4)}$, and
$\mathcal{W}^{(4,1)}$ are generated by the W-currents which satisfy the
conditions \ \eqref{skanzzb} with the pair of
vectors ${\boldsymbol\alpha}_1, \ {\boldsymbol \alpha}_2$ replaced by the corresponding pair
${ \boldsymbol \alpha}_i, \ {\boldsymbol \alpha}_j$, $\ (i,j) = (2,3), (3,4), (4,1)$, respectively.
Of course, all these W-algebras are isomorphic to each other, and differ only in
the way they are embedded in the space of holomorphic fields of the bulk theory
\eqref{action}.

\subsection{Corner-brane}

More interesting conformal boundary conditions are represented by the ``corner-branes'',
the corner surfaces in Fig.\,\ref{fig-corn}. There are four corners in Fig.\,\ref{fig-corn},
suggestively labeled by symbols $(1,2,3)$, $(2,3,4)$, $(3,4,1)$, $(4,1,2)$. Again we
concentrate first on one of them, say the corner $(1,2,3)$. This surface is
described by the equation
\bea\label{corner}
\cos\Big({\textstyle\frac{Z_B}{
\sqrt{n+2}}}\Big)=
 {{\tilde a}}\ \exp\Big(-{\textstyle\frac{X_B}{
\sqrt{n\nu}}}\Big)+ {{\tilde b}}\
\exp\Big(-{\textstyle\frac{Y_B}{
\sqrt{n(1-\nu)}}}\Big)
\, ,
\eea
where
\bea\label{RGtime}
{\tilde a} =\kappa^{\frac{1}{ n\nu}}\, ,\ \ \ \ \  {\tilde b}=
\kappa^{\frac{1}{n(1-\nu)}}\, ,
\eea
and  $\kappa=E_*R$.

Like the hairpin cylinder, the corner-brane boundary condition is equivalent to an
RG fixed point. The RG flow can be ``arrested'' by redefining the RG transformation -
supplementing with the shift $\textstyle{(X,Y,Z)
\to (X + \frac{\delta t}{\sqrt{n\nu}}, Y + \frac{\delta t}
{\sqrt{n(1-\nu) }},  Z)}$. The corresponding linear dilaton has the form
\bea\label{ksai}
\Phi({\bf X}) =-{X\over \sqrt{n\nu}}-{Y\over \sqrt{n(1-\nu)}}
\ .
\eea

In the presence of a dilaton field the one-loop RG flow equations
\eqref{RGfl},\,\eqref{RGflo} are modified as follows \cite{Leigh:1989jq}.
The first of these equations receives an additional term,
\bea\label{RGfldill}
{\bf n}\cdot {\dot {\bf X}}_{B}-
{K\over  1+B_{\rm n}^2}-{\bf n}\cdot{\bf \nabla}\Phi=0\ ,
\eea
whereas $D_j$ in \eqref{RGflo} are replaced by
\bea\label{uuykjsajj}
D_j=({\bf n}\cdot {\dot {\bf X}}_{B})\ B_j+
{1\over 1+B_{\rm n}^2}\
{\partial B_{\rm n}\over \partial \eta^j}-B_{\rm n}\ {\partial\Phi\over
\partial
\eta^j}\ .
\eea
The surface \eqref{ksai} solves the fixed point equations in the presence of
the constant ${\bf B}$ field,
\bea\label{smasjsu}
{\bf B}=\ri\ \big(-\sqrt{1-\nu},\, \sqrt{\nu},\, 0\, \big)\ .
\eea
This can be checked for the one
loop equations,
but it likely holds to all loops.

\subsubsection{\label{secfour23} Corner-brane $W$-algebra}

The corner-brane boundary is conformally invariant. It also very likely to have
certain W-algebra symmetry. In this subsection we will identify the corner-brane
W-algebra, using general arguments and consistency. Beyond the classical limit,
we do not know how to derive the W-algebra directly from the boundary constraint
\eqref{corner}. However, one can {\it define} the corner-brane boundary CFT by the
 W-algebra symmetry, and then check that its semiclassical limit agrees with Eqs.\eqref{sksasaaa},\eqref{corner}.
 This is the strategy we adopt here.

Concentrating again on the corner brane $(1,2,3)$ in Fig.\,\ref{fig-corn}, we observe that
this surface incorporates two hairpin cylinders, $(1,2)$ and $(2,3)$ in Fig.\,\ref{fig-cross},
as the limiting cases. This suggests that the W-algebra associated with
the corner brane $(1,2,3)$ may include holomorphic currents $W_s$ from
$\mathcal{W}^{(1,2)}$, which also belong to $\mathcal{W}^{(2,3)}$. Thus, we define
$\mathcal{W}^{(1,2,3)} = \mathcal{W}^{(1,2)}\,\cap \,\mathcal{W}^{(2,3)}$. In other
words, the W-algebra $\mathcal{W}^{(1,2,3)}$ consists of the currents $W_s$ which
satisfy the condition \eqref{skanzzb} with all three vertex operators
$\exp\big({\boldsymbol \alpha}_j {\bf X}_R\big),\ j=1,2,3$. This W-algebra is not
new. It was introduced in Refs.\cite{fate,Fateev:1996ea}, and further studied in
Ref.\cite{Feigin:2001yq}, where it was named ${\cal W}D(2|1;\alpha)$, with the
parameter $\alpha$ related to our $\nu$ as $\alpha=-\nu^{-1}$. Its Virasoro central
charge is
\bea\label{cw}
c=3+6\ \big({\textstyle{1\over n\nu}+{1\over n(1-\nu)}-{1\over n+2}}\big)\, .
\eea

Likewise, the W-algebras of the other corner branes
in Fig.\,\ref{fig-corn} are defined as the intersections of pairs of
corresponding hairpin W-algebras, e.g., $\mathcal{W}^{(2,3,4)}
=\mathcal{W}^{(2,3)}\,\cap\,\mathcal{W}^{(3,4)}$, etc. Of course, as the
algebras, all these are isomorphic to $\mathcal{W}^{(1,2,3)}$, differing from
it only in the way they are embedded in the space of the chiral fields
of the bulk theory \eqref{action}.

Let us describe here some properties of the algebra ${\cal W}D(2|1;\alpha)$, taking for
definiteness its realization as $\mathcal{W}^{(1,2,3)}$. As usual, the number of independent
holomorphic currents $W_s$ of spin $s$ can be read out of the character of the vacuum representation
of this W-algebra \cite{Feigin:2001yq},
\bea\label{sdlsdj}
\chi_{\rm vac}(q)=1+q^2+q^3+3\, q^4+3\, q^5+8\, q^6+
9\, q^7+19\, q^{8}+25\, q^9+\ldots\ .
\eea
Spin-1 currents are absent, but there is one spin-2 current
\bea\label{energymomaa}
W_2 = - \partial_z  {\bf X}\cdot \partial_z {\bf X}
-
{\boldsymbol \rho}\cdot
\partial_{z}^2{\bf X}
\eea
with
\bea\label{nsaash}
{\boldsymbol \rho}=
\big( {\textstyle{1\over\sqrt{n\nu}}},\,
{\textstyle{1\over\sqrt{n(1-\nu)}}},\,
-{\textstyle{\ri\over\sqrt{n+2}}}\, \big)\ ,
\eea
which generates the Virasoro subalgebra with the above central charge.
Furthermore, there is no truly independent spin-3 currents, since the
only spin-3 field accounted in \eqref{sdlsdj} is the derivative $\pz W_2$.
At spin-4 there are three fields -- two ``descendant'' currents, $\pz^2 W_2$
and $W_{2}^2$, but also one new current $W_4$. By descendants here we understand
the $\pz$ derivatives and composite fields built from of the lower-spin
currents.\footnote{These are
indeed descendants with respect to the W-algebra in the usual CFT sense:
they are obtained from the identity operator by successive applications of
the mode operators of the currents of the lower spins.} Explicit form of $W_4$
(first presented in \cite{Fateev:1996ea})
is somewhat cumbersome, and we relegate
it to  Appendix. It will be important for our arguments below
that it can be written as\footnote{The current\ \eqref{wwwcurrent}\ is not
conformal primary, but can be made a primary by adding certain linear
combination of $W_{2}^2$ and $\pz^2 W_2$; this form can be found
in Ref.\cite{Feigin:2001yq}.}
\bea\label{wwwcurrent}
W_4=W^{({\rm sym})}_4+\partial_z {V}_3\, ,
\eea
where the non-derivative term $W^{({\rm sym})}_4$ (but not $V_3$) is
symmetric with respect to all $180^o$ rotations around the coordinate
axes of the $(X,Y,Z)$ space (equivalently, the simultaneous sign reversals
of any pair of the fields $(X,Y,X)$), and also respects the
symmetries \eqref{jssyay} and
\eqref{jddssyay}. Since these transformations interchange different
corners in Fig.\,\ref{fig-corn}, this symmetry implies that the form of
$W^{({\rm sym})}_4$ is the same for all four realizations of the W-algebra,
associated with the four corners.

It is likely (but not proven, or really verified beyond the spin 5) that
similar structure persists to higher spins. Three-dimensional space of the
spin-5 currents is spanned by the descendants, the $\pz$ derivatives of
the spin-4 currents. The eight-dimensional space of spin-6 currents involves
seven descendants, and one new field, which
again can be written in the form similar to \eqref{wwwcurrent},
\bea\label{wwwcurrentn}
W_6=W^{({\rm sym})}_6+\partial_z {V}_5\ ,
\eea
where $W^{({\rm sym})}_6$ is symmetric under the rotations $(X,Y,Z)\to(X,-Y,-Z)\, ,\, (-X,Y,-Z).$
Going further
up in the spins, we conjecture that there is exactly one independent
current at each even spin, having the form
\bea\label{wwwcrentn}
W_{2k}=W^{({\rm sym})}_{2k}+\partial_z { V}_{2k-1}\ ,
\eea
in which the term $W^{({\rm sym})}_{2k}$ is the same for all four realizations
of the W-algebra associated with the four corners in Fig.\,\ref{fig-corn}.


\subsubsection{Boundary state of the corner-brane}

As usual for conformal boundaries\,\cite{Callan:1988wz, Ishibashi:1988kg, Cardy:1989ir}, the statement of the
${\cal W}D(2|1;\alpha)$-algebra symmetry of the boundary state associated with the
corner-brane can be expressed as the set of equations
\bea\label{wcftinv}
\big[\,
z^s\ W_s(z) -(-1)^s\,
{\bar z}^s \ {\bar W}_s({\bar z})\, \big]_{|z|=R} \
|\, B\, \rangle_{(1,2,3)} = 0\, .
\eea
We have verified this equation only for $s=2$ and $4$, and only in the classical
limit $n\to\infty$.\footnote{The leading term in the $n\to \infty$ asymptotic of
$W_4$ in Eq.\eqref{zzajasjsa} is proportional to $W_{2}^2$; for this term the equation
\eqref{wcftinv} with $s=4$ is a simple consequence of the $s=2$ equation, and thus
does not imply any additional symmetry beyond the conformal invariance. However,
the combination ${\tilde W}_4 = (W_4 + 9\,n^3\nu (1-\nu)\,W_{2}^2)/n^4$ has nontrivial
$n\to \infty$ limit, independent from $W_2$. It is for this current ${\tilde W}_{4}$
that we have checked that $z^4 {\tilde W}_4 - \zb^4 {\bar {\tilde W}}_4$
vanishes in virtue of the classical boundary conditions of the corner-brane.}

It is sensible, however, to take \eqref{wcftinv} as a part of the definition
of the corner-brane boundary. The the boundary state associated with, say, the corner-brane
$(1,2,3)$, will then appear as the combination of the Ishibashi states
 $\mid I_{\bf P}\,\rangle$ of the W-algebra $\mathcal{W}^{(1,2,3)}$ (see e.g.
 \cite{Cardy:1989ir} for the general notion),
\bea\label{boundst}
|\,  B\,  \rangle_{(1,2,3)} = \int\ \rd {\bf P}\
B ({\bf P})\ |\,
I_{\bf P}\, \rangle_{(1,2,3)}\,,
\eea
where $B({\bf P})$ is the vacuum overlap of the corner-brane boundary
state, $B({\bf P})=\langle\, {\bf  P}\, |\, B\, \rangle_{(1,2,3)}$. Unlike the Ishibashi
states, the amplitude $B({\bf P})$ is not determined by W-algebra alone. In principle,
its form is restricted by the requirement of locality of the boundary interaction, but
so far no direct way of solving for this condition is known. Similar problem was
addressed in \cite{Lukyanov:2003nj} for simpler case of the hairpin-brane, where exact form
of the corresponding boundary overlap function was conjectured. Analogous expression
for the corner-brane has the form
\bea\label{bounhair}
&&B (P_1,P_2,P_3) = {g_D^3\over 2\pi}
\ \Big({\kappa\over \alpha^2_1}\Big)^{- \rm{ i}\,  {P_1\over
\alpha_1}}\
\Big({\kappa\over \alpha^2_2}
\Big)^{- \rm{ i}\,  {P_2\over\alpha_2 }}\times
 \\ && \ \ \ \
{\sqrt{\alpha_1\alpha_2}\ \, \Gamma(\, \ri\alpha_1\,P_1\, )\,
\Gamma\big(\, 1 + \ri\, {P_1\over \alpha_1}\, \big)\,
 \Gamma(\, \ri\alpha_2\,P_2\, )\,
 \Gamma\big(\, 1 + \ri\, {P_2\over \alpha_2}\, \big)
\over
\Gamma\big(\, {1\over 2}+
\ri\, {\alpha_1P_1\over 2}+
\ri\, {\alpha_2P_2\over 2}+\ri\, {\alpha_3P_3\over 2}\, \big)
\,
\Gamma\big(\,{1\over 2}+
\ri\, {\alpha_1P_1\over 2}+
\ri\, {\alpha_2P_2\over 2}-\ri\, {\alpha_3P_3\over 2}\, \big)}\, ,
\nonumber
\eea
where
\bea\label{ssjsaut}
\alpha_1= -\sqrt{n\nu}\, ,\ \ \
\alpha_2= -\sqrt{n(1-\nu)}\, ,\ \ \ \alpha_3=-\ri\ \sqrt{n+2}\, .
\eea
This expression passes several simple tests. Thus, it is straightforward to
verify that in the semiclassical limit  \eqref{bounhair} agrees with the result of
direct saddle-point calculation with the corner-brane boundary condition. Also,
note the poles at $P_1=0$ and $P_2=0$ which signify the infinite extent of the
brane in the $X$ and $Y$ directions. The residues at the poles coincide with
the boundary amplitudes of the hairpin-branes of Ref.\cite{Lukyanov:2003nj}, in agreement
with the fact that the corner-brane approaches the hairpin cylinders as in the
limits $X\to\infty$ or $Y\to\infty$. We conjecture that \eqref{bounhair} is exact
boundary amplitude of the corner-brane $(1,2,3)$.

The amplitude \eqref{bounhair} applies to the corner-brane $(1,2,3)$ in
Fig.\,\ref{fig-corn}. The other corner branes are obtained from this one
by simultaneous change of signs of two of the coordinates $(X,Y,Z)$.
Therefore, the boundary amplitudes of the other branes in
Fig.\,\ref{fig-corn} are given by the same expression \eqref{bounhair}
with the signs of two of the components of ${\bf P} = (P_1,P_2,P_3)$
reversed. For example $\langle\,{\bf P}\mid B\,\rangle_{(2,3,4)} =
B(-P_1, -P_2, P_3)$.

\section{\label{secfive2} Integrability of the pillow-brane model}

In this section we will argue that the pillow-brane model is
integrable. The following discussion is closely parallel to that presented in
Ref.\cite{Lukyanov:2003nj} in the context of the ``paperclip model'', therefore we will be
brief.

Let us first remind what we mean by the statement of integrability in this
situation. In the bulk, we are dealing with the theory of free bosons which
is trivially integrable. In particular, the bulk theory has infinite number of
commuting integrals of motion. The relevant integrals of motion look simpler
in terms of the coordinates $(v,{\bar v})=(\tau+\ri\,\sigma,\tau-\ri\,\sigma)$ related
to $(z,\zb)$ via the logarithmic conformal transformation
\bea
v=-\ri\, R\,\log(z/R)\,, \qquad {\bar v}=\ri\, R\,\log(\zb/R)\,,
\eea
which maps the disk $|z|<R$ onto the semi-infinite cylinder $\tau\equiv
\tau+2\pi R$, $\sigma > 0$, with the boundary placed at $\sigma=0$.
The exponential field insertion in \ \eqref{exponent} is equivalent
to the condition that the shifted field ${\bf X} - \textstyle{\ri\,
{\sigma\over R}}\,{\bf P}$ (or \eqref{shift})  is bounded on the
cylinder. 
The space of  holomorphic fields of the bulk theory is spanned
by polynomials $P(v)=P(\partial_v X^j, \partial_{v}^2 X^j, ...)$ of the
components of $\partial_v {\bf X}$ and higher $\partial_v$ derivatives of $X^j$.
Integrating such polynomials over closed contour around the cylinder one
obtains integrals of motion (IM) $\mathbb{I}[P] =\oint\,\frac{ \rm {d} v}{2\pi}\,P(v)$ of the
bulk theory. In the Hamiltonian picture with the coordinate $\sigma$ along the
cylinder taken as the ``time'', the integrals $\mathbb{I}[P]$ are operators
acting in the space \eqref{fullspace}. Let $P_{s+1}(v)$ be a set of polynomials
such that the associated integrals
\bea\label{ims}
\mathbb{I}_s = \oint\,\frac{\rd v}{2\pi}\,P_{s+1}(v)
\eea
all commute, $\big[\,{\mathbb I}_s\,,\,{\mathbb I}_{s'}\,\big] = 0$. We say that
a boundary condition at $\sigma=0$ is consistent with the set
of IM $\{\mathbb{I}_s\}$ if the corresponding boundary state $\mid B\,\rangle$
satisfies the equations \cite{Ghoshal:1993tm}
\bea\label{integrable}
\big(\mathbb{I}_s - {\bar{\mathbb I}}_s\big)\,\mid B\,\rangle =0\,,
\eea
where ${\bar{\mathbb I}}_s = \oint\,\frac{ \rm {d}{\bar v}}{2\pi}\,
{\bar P}_s({\bar v})$ are the corresponding ``left-moving'' IM, obtained
from $\mathbb{I}_s$ by replacing $v \to {\bar v}$.\footnote{Eq.\eqref{integrable}
implies that the differences $P_{s+1}(v)-{\bar P}_{s+1}({\bar v})$, when specified
to the boundary $v={\bar v}=\tau$, reduce to total derivatives $\partial_\tau Q_s(\tau)$
in virtue of the boundary conditions. These equations in turn lead to nontrivial integrals
of motion in the Hamiltonian picture where $\tau$ plays the role of (Matsubara) time.
Indeed, it is easy to see that the quantities $\int_{0}^{\infty}\,\rd\sigma\,
\big(\,P_{s+1}(\tau+\ri\,\sigma)-{\bar P}_{s+1}(\tau-\ri\,\sigma)\,\big) + Q_s(\tau)$ are independent of
$\tau$ \cite{Ghoshal:1993tm}.} A boundary
theory is integrable if it is consistent with a ``maximal'' set of commuting IM.
By definition, any IM $\mathbb{I}[P]$ which commutes with all members of the maximal
set is a linear combination thereof. Generally, a free boson theory admits more then one
maximal set. The maximal commuting set $\{\mathbb{I}_s\}$ is the most important
characteristic of an integrable boundary theory. In this section we identify what
we believe is the commuting set associated with the pillow brane model.

\subsection{ Integrals of motion of the pillow-brane model} \label{secfive1}

As usual, we assume that the subscript $s$ indicates the Lorentz spin of the
IM $\mathbb{I}_s$. First of all, the maximal commuting set $\{\mathbb{I}_s\}$
always include the the spin-1 operator
\bea\label{Ione}
{\mathbb I}_1 = -\oint {\rd v\over 2\pi}\,
 \big(\partial {\bf X}\big)^2\,,
\eea
the light-cone component of the energy-momentum. To figure out what the higher-spin
IM associated with the pillow brane could possibly be, recall that in the
short-scale limit the pillow surface tends to a juxtaposition of four corner
branes. As was explained in Section\,\ref{secfour2}, the individual corner branes are conformal
boundary conditions which enjoy the W-algebra symmetry. Since the IM $\mathbb{I}_s$
of the bulk CFT do not involve any scale, they have to be the elements of all four
W-algebras $\mathcal{W}^{(a,b,c)}$ corresponding to the four corners $(a,b,c) =
(1,2,3), (2,3,4), (3,4,1), (4,1,2)$ in Fig.\,\ref{fig-corn}. Although
$\mathcal{W}^{(a,b,c)}$ are isomorphic to each other as the algebras, they
have different realizations in terms of the chiral fields of \eqref{action}.
The $W$-currents of $\mathcal{W}^{(a,b,c)}$ are defined by the relations
\eqref{skanzzb} with three exponentials, $j=a,b,c$. This suggests that the currents
$P_{s+1}(v)$ in the IM $\{\mathbb{I}_s\}$ must obey some similar relations with all four
exponentials $j=1,2,3,4$. It is highly unlikely that any local current can satisfy
exactly \eqref{skanzzb} with $j=1,2,3,4$. However, the desired IM are integrals
\eqref{ims}, therefore it is sufficient to demand
\bea\label{psdef}
\oint_z\ \rd w\ P_{s+1} (z)\,\,\re^{{\boldsymbol \alpha}_j
\cdot{\bf X}_R}(w)  =\pz F_s\,,
\eea
for $j=1,2,3,4$, where $F_s$ are local fields. The set of currents $P_{s+1}$
defined by this condition is definitely not empty: it is straightforward to verify
that $P_2 = -(\partial_v {\bf X})^2$, and $P_4 = W_{4}^{(\text{sym})}$ given by
Eq.\eqref{zzajasjsa} satisfy \eqref{psdef}. This result was previously obtained
in \cite{Fateev:1995ht} in connection with different model. Moreover, the results of
\cite{Fateev:1995ht} suggest that there is infinite set of currents $P_{s+1}$ with
$s = 1,\,3,\,5,\,\ldots\, 2k-1,\, \ldots$ which satisfy \eqref{psdef}. These are the currents
$W_{2k}^{(\text{sym})},\ k=1,\,2,\, \ldots$ conjectured in Section\,\ref{secfour2}, Eq.\eqref{wwwcrentn}.
Unfortunately, at the moment we do not know how to prove existence of the higher spin
(i.e. beyond $P_2$ and $P_4$) currents with this property. Nonetheless we will
proceed under assumption that an infinite set of currents $P_{2k}, \ k=1,2,3,\,\ldots$
satisfying \eqref{psdef} exists, and moreover that the associates IM
$\mathbb{I}_{2k-1}$ form a maximal commuting set. Then, the boundary state satisfying
\eqref{integrable} has the general form
\bea\label{bstate}
|\,  B\,  \rangle = \int_{\bf P}\, \sum_m \, B_m ({\bf P}) \ |\, m\, ,\,
{\bf P}\,  \rangle \otimes \overline{|\, m\, ,\,  {\bf P}\, \rangle}\,,
\eea
where $|\, m\, ,\,  {\bf P}\, \rangle$ are the orthonormalized\footnote{Here and below we
assume the standard normalization $\langle\,m,{\bf P}\mid m',{\bf P}'\,\rangle =
\delta_{m,m'}\,\delta^{(3)}({\bf P}-{\bf P}')$.}
simultaneous
eigenvectors of the operators ${\mathbb I}_s$ in the space
${\cal F}_{\bf P}$.

Note that the transformations \eqref{jssyay} and \eqref{jddssyay} act by permutations
on the four exponentials $e^{{\boldsymbol \alpha}_j {\bf X}_R},\ j = 1,\,2,\,3,\,4$. Therefore
the currents $P_{s+1}$ defined by \eqref{psdef}, and the IM $\mathbb{I}_s$,  are
expected to be invariant with respect to these transformations (these symmetries
are explicit in the expression \eqref{zzajasjsa} for $P_4$). It is these symmetries that
suggest that Eqs.\eqref{pillow},\,\eqref{sjksus} provide perturbatively exact description of
the pillow-brane -- it is the simplest expression with correct $n\to\infty$ limit which
respects these symmetries.

\subsection{Infrared limit of the pillow-brane model}\label{secsix}

To facilitate the discussion in this section, it is convenient to use formal
interpretation of the boundary state $\mid B\,\rangle$ in terms of the associated
{\it boundary state operator} (see e.g. \cite{Ishibashi:1988kg,Cardy:1989ir}). Isomorphism
between the Fock spaces $\mathcal{F}_{\bf P}$
and ${\bar{\mathcal F}}_{\bf P}$ provides one to one correspondence between the
states in $\mathcal{F}_{\bf P} \,\otimes \,{\bar{\mathcal F}}_{\bf P}$ and operators
in $\mathcal{F}_{\bf P}$. Let $\mathbb{B}$ be the operator corresponding to the boundary
state $\mid B\,\rangle$. Then Eq.\eqref{integrable} is equivalent to the commutativity
\bea\label{commut}
[\, {\mathbb B}\, ,\, {\mathbb  I}_s\, ] = 0\,,
\eea
and Eq.\eqref{bstate} can be written as
\bea
\mathbb{B} = \sum_{m} \,B_m ({\bf P})\,\,
\mid m,{\bf P}\,\rangle\langle\,m, {\bf P}\mid\,.
\eea
Thus, the coefficients $B_m({\bf P})$ are interpreted as the eigenvalues of the
boundary state operator,
\bea
{\mathbb B}\ |\, m\, ,\,  {\bf P}\, \rangle = B_m ({\bf P})\
|\, m\, ,\,  {\bf
P}\, \rangle\,.
\eea
The eigenvalue $B_0 ({\bf P})$ corresponding to the Fock vacuum $
\mid {\bf P}\,\rangle \equiv |\, 0\, ,\,  {\bf P}\,  \rangle$ in ${\cal F}_{\bf P}$ coincides
with the overlap\ \eqref{partit}.

Arguments completely parallel to those given in Ref.\cite{Lukyanov:2003nj} (see Section 6 therein)
for the paperclip model suggest the following large-$R$ expansion of the boundary
state operator of the pillow-brane,
\bea\label{bass}
\log \mathbb{B}\, \asymp\, \log(g_{D}^3) - \sum_{k=0}^{\infty}\,C_k \
E_{*}^{1-2k}\ \ {\mathbb I}_{2k-1}\,
\eea
where $\log(g_{D}^3) = -\frac{3}{4}\,\log 2$ is the boundary entropy  \cite{Affleck:1991tk}
of the infrared
fixed point (the Dirichlet boundary condition\footnote{
The expansion in terms of the local IM appears
in integrable boundary theories which flow down to the ``basic'' boundary fixed
point, the one  which admits  no primary boundary fields but the
identity operator. In those theories
the expansion in ${\mathbb I}_s$ corresponds to expansion of the infrared
effective
action in terms of descendants of the identity. Clearly, the Dirichlet
boundary is of that kind.}, ${\bf X}_B ={\bf 0}$),
and the (asymptotic) series
involves all local IM $\{\mathbb{I}_{2k-1}\}$, with the first one $\mathbb{I}_{-1}$ being the
identity operator by definition. The dimensionless coefficients $C_k$ are yet to
be determined. Even without knowing the coefficients $C_k$, this expansion
has much predictive power. In particular, it yields
the infrared asymptotic expansion of the partition function
\bea\label{ssdjksa}
\log Z(\, {\bf
P}\, |\, \kappa\, )\, \asymp\,
\log({\rm g}_D^3)-\sum_{k=0}^{\infty}C_k\,\,
\kappa^{1-2k}\ I_{2k-1}({\bf P})
\ ,
\eea
where
\bea\label{imexpect}
I_{2k-1}({\bf P}) = R^{2k-1}\,\,\langle\,{\bf P}\mid\mathbb{I}_{2k-1}\mid {\bf P}\,\rangle
\eea
are the dimensionless vacuum expectation values  of the local IM. The expectation values
incorporate all dependence on the components of the zero mode momentum
${\bf P}$. They are polynomials of the degree $2k$ of the
components $(P_1,P_2,P_3)$, which can be obtained by direct computations once
explicit expressions for the currents $P_{2k}$ are known. Thus, from $P_2
=-(\pz {\bf X})^2$ and $P_4 = W_{4}^{(\text{sym})}$ as given by Eq.\eqref{zzajasjsa}
in  Appendix, one finds
\bea\label{sjusyu}
I_{1}({\bf P}) &=& {P^2_1\over 4} + {P^2_2\over 4}+
{P^2_3\over 4} -{1\over 8}\,,
\\
I_{3}({\bf P}) &=&
 \sum_{j=1}^3E_j\, \Big({P_j^4\over 16}-{P_j^2\over 16}+{1\over 192}\Big)+
\sum_{
m\not= j}E_{mj}\, \Big({P_m^2\over 4}-{1\over 24}\Big)
\Big({P_j^2\over 4}-{1\over 24}\Big)+
\nonumber \\ &&{1\over 240}\
\sum_{j=1}^3H_{j}\ ,
\nonumber
\eea
where the numerical coefficients $E_j,\ E_{mj}$ and $H_j$ are given by
\eqref{kasuaiu}.

\section{Proposal for the
overlap amplitude $\langle\, {\bf P}\mid B\,\rangle$}\label{secseven}

In Ref.\cite{Lukyanov:2003nj} exact expression for the boundary amplitude of the paperclip
model was proposed in terms of solutions of certain linear ordinary differential
equation. Here we present similar proposal for the pillow brane model, and
test it against various known properties of the model. Similar constructions
are known in a number of integrable boundary models, where the boundary states
can be related to Baxter's operators \cite{AlZ} (for review see \cite{Dorey:2007zx} and references therein).
The relation of eigenvalues of
Baxter's operators in CFT to ordinary differential equation was originally
proposed in \cite{Dorey:1998pt}.

\subsection{Differential equation}\label{secseven1}

Consider the ordinary second order differential equation
\bea\label{diff}
\Big[\,  -{{\rd^2}\over{\rd x^2}}+V(x)\, \Big]\,  \Psi(x) = 0\ ,
\eea
with
\bea\label{qsaqsksai}
V(x)=\kappa^2\ \re^{-n\nu x}\, \big(1+
\re^x\big)^n-
 {\alpha^2+\beta^2\, {\re^x}\over{1+\re^x}} -
\big(\gamma^2-{\textstyle{1\over 4}}\big)\  { {\re^x}\over{(1+\re^x)^2}}
\ .
\eea
The parameters $\alpha$, $\beta$ and $\gamma$ here will be related
to the components of the momentum ${\bf P} = (P_1,P_2,P_3)$ in
\eqref{exponent},
\bea\label{PQpq}
\alpha={\textstyle{1\over 2}}\ \sqrt{n\nu}\ \, P_1\, ,
\ \ \ \ \beta={\textstyle{1\over 2}}\ \sqrt{n(1-\nu)}\ \, P_2\, ,
\ \ \ \ \gamma={\textstyle{1\over 2}}\  \sqrt{n+2}\  \, P_3\, ,
\eea
and $\kappa$ is assumed to be the same as in \eqref{kappadef}.
Below we always assume that $\kappa$ is real and positive. In the semiclassical
case $n\gg 1$ the parameters $\alpha$, $\beta$, $\gamma$ here are the same as
in\ \eqref{kl}.

Let
$\Psi_{-}(x)$ be the solution of\ \eqref{diff}\ which decays when $x$ goes
to
$-\infty$ along the real axis, and $\Psi_{+}(x)$ be another solution
of\ \eqref{diff},
the one which decays at large positive $x$. We fix normalization
of these two solutions as follows,
\bea\label{psiassminus}
\Psi_{-} &\to&\ \kappa^{-{1\over 2}}\
\re^{\Phi(\nu; x)}
\ \ \ \ \ \ \ \ \ \ \ \ {\rm as}\ \ \ \  x\to-\infty\ ,
\\
\Psi_{+} &\to&\ \kappa^{-{1\over 2}}\ \re^{\Phi(1-\nu; -x)}
\ \ \ \ \ \ \ \ {\rm as}\ \ \ \  x\to+\infty\ ,\nonumber
\eea
where
\bea\label{nahgg}
\Phi(\nu \, |\, x)=
{\textstyle{n\nu x\over 4}}
-{\textstyle {2\kappa\over n\nu}}\ \re^{-{n\nu x\over 2}}\
{}_2F_{1}\big(\, -{\textstyle{n\nu\over 2}},
-{\textstyle{n\over 2}},\, 1-{\textstyle{n\nu\over 2}};-\re^{x}\big)\ .
\eea
Let
\bea\label{wronskian}
W[\Psi_{+},\Psi_{-}]\equiv \Psi_{+}\, {\rd\over \rd x}\,  \Psi_{-} -
\Psi_{-}\, {\rd\over \rd x}\,  \Psi_{+}
\eea
be the Wronskian of these two solutions. Then, our proposal
for the function\ \eqref{partit} is
\bea\label{exactz}
 Z(\, {\bf P}\, |\, \kappa\,) ={{\rm g}^3_D\over 2}\
 \ W[\Psi_{+},\Psi_{-}]\, .
\eea

In this section we present arguments supporting our
proposal \eqref{exactz}. This requires understanding some properties of the solution
of the differential equation \eqref{diff},\,\eqref{qsaqsksai}.

\subsection{Small $\kappa$ }\label{secseven2}

In the analysis below, it will be
convenient to split the potential \ \eqref{qsaqsksai} into two parts,
\bea\label{diffop}
V(x)=V_{-}(x) + V_{+}(x)\, ,
\eea
where
\bea\label{Uminus}
V_{-}(x) =-
 {\alpha^2+\beta^2\, {\re^x}\over{1+\re^x}} -
\big(\gamma^2-{\textstyle{1\over 4}}\big)\  { {\re^x}\over{(1+\re^x)^2}}\, ,
\eea
and
\bea\label{Uplus}
V_{+}(x) =\kappa^2\ \re^{-n\nu x}\, \big(1+
\re^x\big)^n\ .
\eea

At large negative $x$ the term $V_{-}(x)$ approaches the constant $-\alpha^2$,
while $V_{+}(x)$ can be approximated as $\kappa^2\ \re^{-n\nu x}$. At small $\kappa^2$,
the accuracy of the
last approximation is understood after making the change of the variable
$x=x_0 - {\textstyle {2\over n\nu}}\, y$,
where $x_0 = {2\over n\nu}\,\log\big({\textstyle{2\kappa\over n\nu}}\big)$;
Eq.\,\eqref{diff}\ then takes the form
\bea\label{difff}
\Big[-\frac{\rd^2}{\rd y^2} -
\Big(\frac{2\alpha}{ n\nu}\Big)^2
+ \re^{2y} + \delta V(y)\, \Big]\,\Psi
=0\,,
\eea
where $\delta V\sim\kappa^{2\over n\nu}$ as $\kappa\to 0$.
Therefore for $1\ll -x$ we have
\bea\label{sjuyy}
\Psi_{-}(x) = {2\over\sqrt{\pi n\nu}}\
 K_{ {\textstyle
 {2{\rm i}\, \alpha\over n\nu} }} \big( {\textstyle {{2\kappa}\over
  n\nu}\,\re^{-{n\nu x
\over 2}}}\big) + O\big(\kappa^{2\over n\nu}\big)\, .
\eea
where $K_{\mu}(z)$ is the Macdonald function. The normalization in
Eq.\eqref{sjuyy}\ is chosen to agree with the asymptotic form\ \eqref{psiassminus}.
In the domain
\bea\label{sksunag}
1\ll -x
\ll {\textstyle{1\over n\nu}}\
\log\big({\textstyle{1\over \kappa^{2}}}\big)\, ,
\eea
the solution $\Psi_{-}$\ \eqref{sjuyy}\ becomes a combination
of two plane waves,
\bea\label{waves}
\Psi_{-}(x) = D_{\nu}(\, -\alpha\,)\
 \re^{+{\rm i} \alpha\,x} + D_{\nu}(\, \alpha\,)
\ \re^{-{\rm i} \alpha \,x}\,,
\eea
with
\bea\label{sjdsudu}
D_{\nu}(\, \alpha\,)=
{1\over\sqrt{\pi n\nu}}\ \,
\Big({\kappa\over
  n\nu}\Big)^{
 {2{\rm i}\, \alpha\over n\nu}}
\ \Gamma\Big(  - {2{\rm i} \alpha\over n\nu}\Big)
\ \Big[\, 1+O\big(\kappa^{2\over n\nu}\big)\, \Big]\ .
\eea

On the other hand, when $\kappa$ goes to zero, and
\bea\label{plateau}
-{\textstyle{1\over n\nu}}\
 \log\big({\textstyle{1\over \kappa^{2}}}\big) \ll x
\ll {\textstyle{1\over n(1-\nu)}}\
\log\big({\textstyle{1\over \kappa^{2}}}\big)\, ,
\eea
the term $V_+$\ in \ \eqref{diffop} is negligible. In this domain
Eq.\eqref{diff}\ reduces to the
Riemann differential equation, and its solution $\Psi_-$
specified by the asymptotic behavior\ \eqref{waves}\
has the form
\bea\label{nwaves}
\Psi_{-}(x) = D_{\nu}(\, -\alpha\,)\
Q_{-\alpha,\beta,\gamma}\big(\re^x\big) +
D_{\nu}(\,\alpha\,)
\ Q_{\alpha,\beta,\gamma}\big(\re^x\big)\, ,
\eea
where $Q_{\alpha,\beta,\gamma}(w)$ is the hypergeometric function
\ \eqref{nshshaa}. The solution $\Psi_+$ can be obtained
from $\Psi_-$ by substitution $\alpha\leftrightarrow\beta$,
$\nu\to 1-\nu$ and $x\to -x$, i.e.,
\bea\label{naawaves}
\Psi_{+}(x) = D_{1-\nu}( \,-\beta\,)\ Q_{-\beta,\alpha,\gamma}
\big(\re^{-x}) +
D_{1-\nu}( \,\beta\, )
\ Q_{\beta,\alpha,\gamma}\big(\re^{-x})\, .
\eea
Using
\bea\label{sjsujau}
W\big[\,Q_{\beta,\alpha,\gamma}\big(\re^{-x})\, ,
\,  Q_{\alpha,\beta,\gamma}\big(\re^{x})\, \big]={\Gamma(1-2\ri\alpha)
\, \Gamma(1-2\ri\beta)\over
\Gamma(\, {1\over 2}-\ri\alpha-\ri\beta-\gamma\, )\,
\Gamma(\, {1\over 2}-\ri\alpha-\ri\beta+\gamma\, )}\ .
\eea
we find the limiting $\kappa\to 0$ form of \eqref{exactz},
\bea\label{ssusu}
\frac{g_{D}^3}{2}\ W[\Psi_+,\Psi_-]\ \to\ \sum_{\varepsilon,\varepsilon'=\pm 1}
\,B(\varepsilon\,P_1, \varepsilon'\,P_2, \varepsilon \varepsilon'\,
P_3)\,,
\eea
where $B(P_1,P_2,P_3)$ is exactly the boundary-state amplitude \eqref{bounhair} of the corner-brane.
This is expected form of the
UV limit $\kappa\to 0$ of the boundary amplitude of the pillow-brane model.
Indeed, as was discussed in Section\ \ref{secfour2}, in this limit the pillow surface
becomes infinitely wide in the $X$ and $Y$ directions, and its shape near
the round ends is well described by the corner-branes in Fig.\,\ref{fig-corn}.
If the components $P_1, P_2$ of the momentum have nonzero imaginary part, the
exponential insertion in the functional integral \eqref{path} pulls the field
towards one of the corners. This is why in the $\kappa\to 0$ limit
$Z({\bf P}|\kappa)$ must reduce to the corner brane amplitude. Which of the
corners dominate depends on the signs of $\Im m\, P_1$ and $\Im m\, P_2$,
and the effect is expected to become more prominent at large $n$, where
the classical configuration dominates. This $\kappa\to 0$ limiting behavior
is in full agreement with \eqref{ssusu}, where the factor
$\kappa^{ {{\rm i} \varepsilon\,P_1\over \sqrt{n\nu}}+ {{\rm i} \varepsilon'\,
P_2\over \sqrt{n(1-\nu)}}}$ in $B(\varepsilon\,P_1,
\varepsilon'\,P_2, \varepsilon \varepsilon'\,P_3)$ makes one of the terms
in the sum dominate at nonzero $\Im m\, P_1,\ \Im m\, P_2$.

\subsection{Semiclassical domain $1 \ll n$, $\kappa \ll 1$}\label{secseven3}

It is possible to show that corrections to the $\kappa\to 0$ limiting
form \eqref{ssusu} are expanded in powers of $\kappa^{2\over n\nu}$ and
$\kappa^{2\over n(1-\nu)}$. When $\kappa$ is small but $n$ is large, so that
$\kappa^{2\over n} \sim 1$, all terms in this expansion are to be collected.
Let us assume that the roots $w_1$ and $w_2$
of the equation \eqref{sjksus}\ are not too
close to each other. This regime corresponds to the
semiclassical domain of the paperclip model
considered in Section\ \ref{secthree2}.

First, let us consider the case when $\alpha,\ \beta$ and
$\gamma$ in \eqref{qsaqsksai} are of the order of 1, and the semiclassical
pillow-brane amplitude is given by Eq.\eqref{hssy}.
Under these conditions the term
$V_{+}(x)$ in
the potential has the effect of rigid walls at
some points $x_1,\ x_2$ ($x_1 < x_2$). That is, for
$x-x_1\gg \textstyle{1\over n}$ and
$x_2 - x \gg \textstyle{1\over n}$, the potential
$V_{+}(x)$ is negligible, but outside the segment $(x_1,\,x_2)$
it grows fast, so that the solution $\Psi_{-}(x)\ (\Psi_{+}(x))$ essentially
vanishes at $x<x_1 \ (x>x_2)$.
It is possible to show that when $1 \gg \,x-x_1\,\gg\, 1/n$,
the solution $\Psi_{-}(x)$ is well approximated by a linear function,
\bea\label{linear}
\Psi_{-}(x)  \approx \chi_1\ (x - x_1)\, .
\eea
Similarly, when $x$ is bellow but close to $x_2$, the solution
$\Psi_{+}(x)$ behaves as
\bea\label{linear1}
\Psi_{+}(x)  \approx \chi_2\ (x_2 - x)\, .
\eea
The positions $x_{1}, x_2$ of the
walls and the slopes $\chi_{1,2}$ depend on $\kappa$,
\bea\label{ksiuim}
\chi_{1,2}=\sqrt{
{n\over \pi}\ {\big|\,\nu-(1-\nu)w_{1,2}\, \big|\over  (1+w_{1,2})}}\
\ \Big(1+
O\big({\textstyle{1\over n}}\big)\, \Big)\ ,
\eea
and
\bea\label{jsysuy}
x_{1,2}=\log (w_{1,2})+O\big({\textstyle{1\over n}}\big)\ ,
\eea
where $w_{1,2}$ are two roots of the equation identical to
the pillow-brane RG flow equation \eqref{sjksus}. We will derive these
equations later in this section. Within the segment $(x_1,\,x_2)$ the
term $V_{+}(x)$ can be neglected, and \eqref{diff} reduces to
hypergeometric equation. Its solutions $\Psi_{\pm}(x)$ in this segment are
uniquely determined by the corresponding ``initial conditions'' \eqref{linear}
and \eqref{linear1}. Thus, for instance
\bea\label{psimm}
\Psi_{-}(x) = \frac{\chi_1}{2\ri\alpha}\,\big(\, Q_{\alpha, \beta, \gamma}(w_1)\,
Q_{-\alpha,\beta,\gamma}(e^x)-Q_{-\alpha, \beta, \gamma}(w_1)\,
Q_{\alpha,\beta,\gamma}(e^x)\,\big)\, , \  \ \ \ x_1 < x < x_2\,.
\eea
To compute the Wronskian in \eqref{exactz}, take $x$ close to
the right wall $x_1$, where both \eqref{linear} and \eqref{psimm} are valid.
Then
\bea\label{asjkxhsj}
&&W[\Psi_+,\, \Psi_-]\approx \chi_2\, \Psi_-(x_2)\,.
\eea
With \eqref{psimm} and \eqref{ksiuim} this yields exactly Eq.\eqref{hssy}.
(Of course, one can make similar analysis in the vicinity of the left wall
$x_2$, which leads to the same result.)

Next, consider the case $P_1,\, P_2,\, P_3 \sim 1$. The term
$V_{+}(x)$ still dominates outside the segment $(x_1,\,x_2)$, but can be
neglected inside it, provided $ x-x_1 \gg {\textstyle{1\over n}}$ and
$ x_2-x \gg {\textstyle{1\over n}}$.
Take $x$ in the vicinity of the left wall, i.e., in the domain
$|x-x_{1} | \ll 1$.
Here the term $V_{-}(x)$ can be replaced by its value at $x_1$,
\bea\label{sbsgtyt}
V_{-}(x)\approx V_-(x_1)  =-{\pi^2\over 4 n}\ \chi^4_1\ \Pi^2_1\ ,
\eea
where $\chi_1$ is given by\ \eqref{ksiuim} and $\Pi_1$ is exactly the
expression \eqref{nash}.  On the other hand, in this  domain the
term $V_+(x)$ behaves as the exponential
\bea\label{snsja}
V_{+}(x)\approx  \pi^2\chi_1^4\ \re^{-2y}\, ,\ \ \ \ \ y=
{\pi\chi_1^2\over 2}\ (x-x_1)\ .
\eea
Therefore, at $x$ close to $x_1$ we have
\bea\label{psiwall}
\Psi_{-}(x)  \approx { {2\over\pi\chi_1}}\
\ K_{{ \rm{i}} \Pi_1\over \sqrt{n}} \big(2\,\re^{-y}\big)\ ,
\eea
where the normalization factor is fixed by matching the asymptotic form
\eqref{psiassminus}.
For small ${\Pi_1/\sqrt{n}}$ and $y \gg 1$ \eqref{psiwall}
reduces to a linear function -- this is how Eq.\eqref{linear}
was obtained. The  solution \eqref{psiwall}\ should be matched to
\eqref{nwaves}\ in the domain
${\textstyle{1\over n}} \ll x-x_1 \ll 1$, where
both approximations  are valid, and then continued to
the right wall. Likewise, in the vicinity of the right wall, where
$|x-x_{2} | \ll 1$, the solution $\Psi_{+}$ is approximated as
\bea\label{psiwalla}
\Psi_{+}(x)  \approx { {2\over\pi\chi_2}}\
\ K_{{ \rm{i}} \Pi_2\over \sqrt{n}} \big(2\,\re^{{\tilde y}}\big)\, ,
\ \ \ \ \ {\tilde y}=
{\pi\chi_2^2\over 2}\ (x-x_2)\ ,
\eea
where $\Pi_2$ and $\chi_2$ are given by\ \eqref{nash}\ and
\eqref{ksiuim}\ respectively. Then
the Wronskian in \eqref{exactz} can be evaluated in the domain
${\textstyle{1\over n}} \ll x_2-x \ll 1$;
the result is exactly \eqref{kjsjsu} with $B_\text{class}$ replaced by
${\tilde B}_\text{class}$, Eq.\eqref{shsdy}.

\subsection{Large $\kappa$ }\label{secseven4}

At large $\kappa$ WKB approximation for the solutions of \eqref{diff} is valid.
Standard calculations within the WKB expansion \cite{Landau}  yield for the
Wronskian \eqref{wronskian}
\bea\label{wkbu}
\log W &=&\log(2)  +{\Gamma(-{\alpha_1^2\over 2})\Gamma(-{\alpha_2^2\over 2})
\over
\Gamma(1+{\alpha_3^2\over 2})}\ \kappa+\\ &&
\int_{-\infty}^{\infty}\rd x\,\bigg\{ \kappa\  \big(
{\cal P}(x)-{\cal P}_0 (x) \big) +
{1\over{8\kappa}}\,{{({\cal P}'(x))^2}
\over{{\cal P}^3(x)}} +\ldots  \bigg\}\ ,\nonumber
\eea
where
${\cal P}(x)=\kappa^{-1}\, \sqrt{V(x)}$ and
we use the notations\ \eqref{ssjsaut}. The
term with ${\cal P}_0 (x) =\re^{-{n\nu x\over 2}}\, \big(1 + \re^x\big)^{n\over 2}$
is subtracted in order to take into account the asymptotic conditions
\eqref{psiassminus}. With the explicit form \eqref{diffop} of the potential,
Eq.\eqref{qsaqsksai} yields asymptotic expansion of the
partition function\ \eqref{ssdjksa} which has the form \ \eqref{ssdjksa} with
\bea\label{maaahh}
C_k= {{\Gamma(k-{1\over 2})}\over{\sqrt{\pi}\, (2k-1)\, 2^{3(k-1)}}}\
{\Gamma\big(1+\alpha_1^2
(k-{1\over 2})\, \big)\, \Gamma\big(1+\alpha_2^2 (k-{1\over 2})\big)\over
\Gamma\big(-\alpha_3^2(k-{1\over 2})\big)}\ ,
\eea
and and  $I_{2k-1}({\bf P})$ being certain polynomials of the variables $P_1^2,\ P_2^2$
and $P_3^2$ (related to $\alpha, \beta, \gamma$ as in \eqref{PQpq}) of the degree $2k$.
The highest-order terms in these polynomials are determined by the first
term in the integrand in\  \eqref{wkbu},
\bea\label{ikhighest}
&&I_{2k-1}({\bf P}) ={ (-2)^{k-1} \over (2k-1)^2
\, (\alpha_1\alpha_2\alpha_3)^2}\,
\sum_{i+j+l=k} (\alpha_1\,P_1)^{2i}\
(\alpha_2\, P_2)^{2j}\ (\alpha_3\,P_3)^{2l}\times
\nonumber\\ && \  \
{
\big(\alpha_1^2\, (k-{\textstyle{1\over 2}})\big)_{k-i}
\big(\alpha_2^2\, (k-{\textstyle{1\over 2}})\big)_{k-j}\,
\big(\alpha_3^2\, (k-{\textstyle{1\over 2}})\big)_{k-l}\over
i!\, j!\, l!}\
+ \ldots \ \ \ \ \ \ (k\geq 0)\, ,
\eea
where we use the notation
\bea\label{slsdo}
(x)_j={\Gamma(x+j)\over \Gamma(x)}\ .
\eea
This expression is in perfect agreement with the highest-order
terms of the polynomials \eqref{sjusyu} obtained directly from
the lowest spin IM. Moreover, it is straightforward to generate the
full polynomials $I_{2k-1}({\bf P})$ evaluating the
integral\ \eqref{wkbu}\ order by order in $\kappa^{-2}$. This
calculation reproduces the eigenvalues \eqref{sjusyu} in all details.
This seems to be highly nontrivial test of our proposal. It would be
interesting to find higher spin representatives of the commuting
set $\{\mathbb{I}_{2k-1}\}$, and compare their vacuum eigenvalues
with the higher polynomials $I_{2k-1}({\bf P})$ in the WKB expansion
\eqref{wronskian}. Note that our proposal predicts exact values of
the coefficients $C_k$ in \eqref{ssdjksa}.

\subsection{$U(1)$-invariant limit}\label{secseven5}

As was mentioned in
Introduction, in the limit $n \to \infty, \ \nu\to 0$ with the
parameters $\lambda=n\nu$ and
${\bar \kappa}^2=\kappa^2\, \nu^{-\lambda}\,
\lambda^{-2}$, the pillow \eqref{pillow} becomes a surface of
revolution \eqref{nshsh}. It is
interesting to look at the differential equation
\eqref{diff} in this limit.

If $n$ goes to $\infty$ while $\kappa$ remains fixed,
the term $V_{+}(x)$ in
the potential\ \eqref{diffop}\ becomes
infinite at any finite $x$. Interesting limit
is obtained by making first the shift of $x$,
\bea\label{logn}
x =y +\log(\nu) \, ,
\eea
so that $V_{+}(x) = \kappa^2\nu^{-\lambda}\,\re^{-\lambda y}\, \big(1+\nu\,
\re^{y}\big)^n$.
Then the
limit $n\to\infty$ brings the equation \eqref{diff} to the form
\bea\label{circdiff}
\Big[ -{ {{{\rm d}^2}\over{{\rm d} y^2}} }
-\frac{\lambda}{ 4}\
\big(\, P_{\parallel}^2+P_{\perp}^2\ \re^{  y}\, \big)
+{\bar \kappa}^2\,\exp\big(\, \lambda\, (\re^{y}-y)\, \big)\, \Big]
\, \Psi(y) = 0\,,
\eea
where $P_{\parallel}=P_1,\ P_{\perp}^2=P_2^2+P_3^2$.
Thus, our proposal\ \eqref{exactz}\ applies
directly to the $U(1)$-invariant brane  \eqref{nshsh}, with
$W[\Psi_{+},\Psi_{-}]$ defined as the Wronskian of two solutions of the
differential equation\ \eqref{circdiff}\ specified by the asymptotic
conditions
\bea\label{minusas}
\Psi_{-}(y) \to {\bar\kappa}^{-{1\over 2}}
\exp\Big(\, {\textstyle{\lambda \over 4}}\,y-2\, {\bar\kappa}\, \re^{-{\lambda
\over 2}\, y}\ G\big(\, {\textstyle{\lambda\over 2}}\ \re^{y}\, \big)
\, \Big) \ \  \
{\rm as}\ \ \   y \to -\infty\, ,
\eea
and
\bea\label{plusas}
\Psi_{+}(y)\to {\bar\kappa}^{-{1\over 2}}\,
\exp\Big(\, {\textstyle\frac{\lambda}{ 4}}\, \big(\, y-\re^{y}\, \big)+
2\, {\bar\kappa}\, \re^{-{\lambda
\over 2}\, y}\ G\big(\, {\textstyle{\lambda\over 2}}\ \re^{y}\, \big)\, \Big)
 \ \ {\rm as}\  \ \  y \to +\infty\, .
\eea
Here $G(y)$
is the confluent hypergeometric function,
\bea
\label{sjsusu}
G(z)={}_1F_{1}\big(
-{\textstyle{\lambda\over 2}},\,  1-{\textstyle{\lambda\over 2}}\, ;\, z\, \big)\ .
\eea

Finally, if one sets
\bea\label{mdsjuju}
y={v\over \sqrt{\lambda}}\ ,
\eea
and takes the limit $\lambda\to \infty$ with fixed $z$ and
${\tilde \kappa}^2={\bar \kappa}^2\ \lambda\, \re^{\lambda}$, the differential
equation\ \eqref{circdiff}\ reduces to
\bea\label{caaircdiff}
\Big[ -{ {{{\rm d}^2}\over{{\rm d} v^2}} }
-{{\bf P}^2\over 4}\,
+{\tilde \kappa}^2\ \re^{{v^2/ 2}}\ \Big]
\, \Psi(z) = 0\ .
\eea
This is exactly the differential equation proposed in Ref.\cite{Lukyanov:2003rt}
in relation to the spherical-brane model.

\section*{Acknowledgments}

Many key points in this work have emerged as the development of ideas of
$\boxed{\rm Alexei\ Zamolodchikov}$ and we were privileged to discuss them with him.

\bigskip
\noindent
We are grateful to
Vladimir Bazhanov and  Vladimir Fateev
for their interest to this work, and to Dmitri Belov for help with
graphic software.
SL also acknowledges discussions with Dmitri Belov, Greg Moore,
Nikita Nekrasov and Alexei Tsvelik.

\bigskip

\noindent
This research is supported
in part by DOE grant $\#$DE-FG02-96 ER 40959.
SL also acknowledges support from Institute
for Strongly Correlated and Complex Systems at BNL
where the part of this work was done in March 2003.

\bigskip
\bigskip

\appendix

\section{ The current $W_{4}$ in Eq.\eqref{wwwcurrent}}

Here we present explicit expression for the current $W_{4}$ in Section\,\ref{secfour2}.
The two terms in \eqref{wwwcurrent} involve
\bea\label{zzajasjsa}
W^{({\rm sym})}_4&=&\sum_{j=1}^3E_j\ \big(\partial_z X^j\big)^4+
\sum_{
m\not=j}E_{mj}\ \big(\partial_z X^m\big)^2\big(\partial_z X^j\big)^2
\nonumber \\ &+&
 \sum_{j\not=k\not=m} K_j\
\partial^2_z X^j\, \partial_z X^k\, \partial_z X^m+
\sum_{j=1}^3H_{j}\ \big(\partial^2_z X^j\big)^2
\eea
and
\bea\label{skssijm}
{ V}_3&=&\sum_{j=1}^3K'_j\ \big(\partial_z X^j\big)^3+
\sum_{m\not=j
}K'_{mj}\ \partial_z X^m\big(\partial_z X^j\big)^2\\ &+&
\sum_{j=1}^3H'_j\ \partial^2_z X^j\partial_z X^j+
\sum_{m\not=j}
H'_{mj}\ \partial^2_z X^m\partial_z X^j+\sum_{j=1}^3F'_j\
\partial^3_z X^j\, .
\nonumber
\eea
In these equations $(X^1,\, X^2,\, X^3)$ stand for $(X,\, Y,\, Z)$,
and the coefficients are expressed through three numbers\ \eqref{ssjsaut}
as follows,
\bea\label{kasuaiu}
&&E_{j}=-\alpha_j^2\ (3\alpha^2_m+2)(3\alpha^2_k+2)
\, ,\nonumber\\
&&E_{mj}=-3\, \alpha_m^2\alpha_j^2\ (3\alpha_k^2+2)\, ,
\\
&&K_j=-2\ \alpha_1\alpha_2\alpha_3
\, \ (3\alpha^2_j+2)
\, ,\nonumber\\
&&H_j=8-\alpha_j^4-9\ (\alpha^2_1\alpha^2_2+\alpha^2_2\alpha^2_3+
\alpha^2_3\alpha^2_1)-
15\ \alpha^2_1\alpha^2_2\alpha^2_3
\, ,\nonumber
\eea
and
\bea\label{kasuaiur}
&&K'_{j}={\textstyle {2\over 3}}
\  \alpha_j\ (3\alpha_m^2+2)(3\alpha_k^2+2)\, ,\nonumber\\
&&K'_{mj}=2\ \alpha_m\alpha_j^2\ (3\alpha_k^2+2)\, ,
\nonumber\\
&&H'_j=\alpha_j^2\ (2\alpha_m^2+2\alpha_k^2+5\, \alpha_m^2\alpha_k^2)\\
\nonumber
&&H_{mj}'=2\, \alpha_m\alpha_j\ (\alpha_m^2-\alpha_k^2)
\\
&&F'_j=-{\textstyle {1\over 3}}\  \alpha_j^3\, ,\nonumber
\eea
In these equations $(j,\,k,\,m)$ represent any permutation of the numbers
$(1,2,3)$.

\end{document}